\documentclass[twocolumn]{aastex631}
\usepackage{bm}
\usepackage{graphicx}
\usepackage{booktabs} 
\usepackage{amsmath} 
\usepackage{xcolor}

\begin{document}

\title{\textbf{\Large{Ring Seismology of the Ice Giants Uranus and Neptune \\}}}

\author[0000-0001-7522-7806]{Joseph A. A'Hearn}
\affiliation{Department of Physics, University of Idaho, Moscow, Idaho, USA}

\author[0000-0002-8592-0812]{Matthew M. Hedman}
\affiliation{Department of Physics, University of Idaho, Moscow, Idaho, USA}

\author[0000-0002-4940-9929]{Christopher R. Mankovich}
\affiliation{California Institute of Technology, Pasadena, California, USA}

\author[0000-0002-3250-2211]{Hima Aramona}
\affiliation{Reed College, Portland, Oregon, USA}

\author[0000-0002-5251-2943]{Mark S. Marley}
\affiliation{Department of Planetary Sciences and Lunar and Planetary Laboratory, University of Arizona, Tucson, Arizona, USA}

\correspondingauthor{Joseph A'Hearn}
\email{josephahearn3@gmail.com}

\begin{abstract} 
We assess the prospect of using ring seismology to probe the interiors of the ice giants Uranus and Neptune. 
We do this by calculating normal mode spectra for different interior models of Uranus and Neptune using the stellar oscillation code {\tt GYRE}. 
These spectra provide predictions of where in these planets' ring systems the effects of interior oscillations might be detected. 
We find that f-mode resonances with azimuthal order $m=2$ or $7 \leq m \leq 19$ fall among the inner rings (6, 5, 4, $\alpha$, and $\beta$) of Uranus, while f-mode resonances with $2 \leq m \leq 12$ fall in the tenuous $\zeta$ ring region. In addition, f-mode resonances with $m=2$ or $6 \leq m \leq 13$ may give azimuthal structure to Neptune's tenuous Galle ring. 
We also find that g-mode resonances may fall in the middle to outer rings of these planets. 
Although an orbiter is most likely required to confirm the association between any waves in the rings and planetary normal modes, 
the diversity of normal mode spectra implies that identification of just one or two modes in the rings of Uranus or Neptune would eliminate a variety of interior models, and thus aid in the interpretation of Voyager observations and future spacecraft measurements. 

\vspace{0.5in}
\end{abstract} 

\section{Introduction \label{intro}}

Three decades after the Voyager flybys of Uranus and Neptune, 
our knowledge of their internal structure and composition is still quite limited \citep{Helled20b}. 
The ice giants Uranus and Neptune represent a distinct class of planets with radii between those of terrestrial worlds like Earth and Venus and gas giants like Jupiter and Saturn. 
The recent discovery that Neptune-mass exoplanets are common \citep{Suzuki16} motivates the exploration of the ice giants in our own solar system.

The deep interiors of the ice giants Uranus and Neptune are of special interest because their internal structure and composition are distinct from the gas giants Jupiter and Saturn. 
Their measured densities suggest that Uranus and Neptune have substantial amounts of carbon, nitrogen, oxygen, sulfur, and silicon, which form compounds like water, ammonia, methane, and silicate minerals \citep{Podolak19}.
At the high pressures and temperatures of planetary interiors, these ``rocks" and ``ices" display interesting properties. 
Water, for example, can enter a superionic state where the oxygen atoms become a lattice and the hydrogen nuclei are free to move, which may play a role in explaining  the non-axisymmetric non-dipolar magnetic fields of Uranus and Neptune \citep{Cavazzoni99, Wilson13}. 
Although numerical calculations and laboratory experiments are providing better constraints on these exotic phases of these materials 
\citep{Knudson12, Bethkenhagen17, Millot18, Millot19, French19M},
and methods to constrain interiors through high-precision gravity measurements are being developed \citep{Movshovitz22},
there remain many uncertainties regarding important parameters such as the mixing properties of the various compounds. In addition, both the overall breakdown and spatial distributions of hydrogen, helium, water, ammonia, methane, and silicates within the planets are poorly constrained. 

While the interiors of Uranus and Neptune remain largely hidden, their ring systems are available for easier observation. 
Recent analyses of Saturn's rings have demonstrated that certain ring features  are likely generated by resonances with normal modes  inside the planet \citep{Hedman13, Hedman14, French16, French19, Hedman19, French21, Hedman22}, confirming predictions made by \citet{Marley93} and providing new insights into that planet's interior. Furthermore, \citet{Marley88} performed preliminary calculations and found that resonances with a few Uranian normal modes could potentially  fall close to some of the Uranian rings
(see Table \ref{Marley_predictions} in the Appendices).  
This paper therefore seeks to build upon that work and establish which planetary normal modes are most likely to fall close to the rings of Uranus and Neptune and thus are the most promising candidates for performing ring seismology at the ice giants.

In Section \ref{bg}, we provide a brief overview of ring seismology and summarize the current state of knowledge about the ring systems.
In Section \ref{meth}, we describe the interior models that we use, and then we explain how we analyze the gravitational potential and calculate the resonance locations.
In Section \ref{res}, we report the locations of these resonances and compare them with relevant structures in the ring systems of the two planets. 
In Section \ref{disc}, we discuss which of the planetary normal mode resonances are most likely to be detectable in the rings, with remarks on several individual rings of interest. Complete tables of mode frequencies and associated resonance locations are provided in the Appendices. 

\section{Background \label{bg}}

Before describing our methods for computing normal mode resonance locations, we first provide a brief overview of giant planet seismology in Subsection \ref{seis_intro} and the relevant features in the rings of Uranus and Neptune in Subsection \ref{rings_intro}.

\subsection{Overview of Ring Seismology \label{seis_intro}}

Planetary oscillations can be decomposed into a set of normal modes, each of which oscillates at a frequency that depends on the planet's profiles of density, adiabatic sound speed, and rotation frequency. Familiar families of oscillation modes include g-modes, whose restoring force is buoyancy, and p-modes, whose restoring force is pressure.
Three numbers characterize planetary normal-mode oscillations: the number of radial nodes $n$,
the spherical harmonic degree $\ell$, and the azimuthal wavenumber $m$.
In this paper, we adopt the convention that $m > 0$ correspond to prograde modes, i.e. modes that propagate in the same direction as the planet's rotation. 
While the mode amplitude spectrum is generally unknown, for the simplest assumption of energy equipartition, the strongest perturbations in the planet's gravitational field are produced by the fundamental modes ($n=0$; f-modes), which can arise from a combination of predominantly gravity but also pressure as the restoring force \citep{Unno79}. 
Saturn's f-modes, for example, produce the most obvious features in its rings 
\citep{Hedman13, Hedman14, French16, French19, French20, Hedman19}, as predicted by \citet{Marley93}.
For this reason, we will primarily consider the f-modes here. 

Because the oscillation frequencies depend on the planet's density profile, 
measurements of these frequencies can probe the planet's internal structure.
Efforts to detect oscillations with visible photometry, which were apparently successful for Jupiter \citep{Gaulme11},
have not yet been successful for the ice giants \citep{Rowe17, Gaulme17}.
\citet{Friedson20} found that for reasonable amplitudes, detection of pressure or temperature variations due to ice giant normal modes is not as promising as the prospect of detecting their gravitational influence on an orbiting spacecraft. 

Fortunately, we can potentially also detect planetary normal modes by treating the ring material that orbits the planet as a seismograph. 
Any even $\ell-m$ mode is symmetric about the equator and would generate a Lindblad resonance in ring material, which excites density waves; whereas any odd $\ell-m$ mode is antisymmetric about the equator and would generate a vertical resonance, which excites bending waves.
The $\ell=m$ modes are the modes that are expected to be the most easily observable at their Lindblad resonance locations because the amplitude of the gravitational perturbation in the ring plane suffers no destructive interference due to latitudinal variations in the phase of the planetary oscillation.
Because Saturn's rings are the largest in our solar system and the dataset on them was the most extensive, they were naturally the first target for ring seismology. 
Voyager images and radio occultation profiles of Saturn's rings revealed spiral density waves and bending waves. 
Some of these waves could be explained in terms of resonances with Saturn's moons, 
but other waves in the C ring were far from any known satellite resonance \citep{Rosen91}. 
Meanwhile, \citet{Marley91} and \citet{Marley93}, building upon ideas from \citet{Stevenson82}, 
showed that  certain normal modes in Saturn's interior could cause gravitational perturbations in the C ring,
and proposed potential correlations between some of the waves and specific planetary f-modes. 
Once the Cassini mission arrived at Saturn, new waves were detected \citep{Baillie11} 
and several features were confirmed to be generated by resonances with planetary oscillations and asymmetries
\citep{Hedman13, Hedman14, French16, French19, Hedman19, French21, Hedman22}. 
These studies yielded the azimuthal wavenumber and the precise frequency for a set of planetary normal modes
that provide evidence for a stably stratified layer within the planet \citep{Fuller14}, 
an estimate for Saturn's bulk rotation rate \citep{Mankovich19},
evidence for a diffuse core \citep{Mankovich21},
and constraints on differential rotation \citep{Dewberry21}. 

\subsection{The Rings and Inner Moons of Uranus and Neptune \label{rings_intro}}

Table \ref{inner_moons} displays the semi-major axes and eccentricities of the thirteen innermost Uranian moons and the seven innermost Neptunian moons. 
Table \ref{rings} shows the semi-major axes and widths of the inner Uranian and Neptunian rings. 

The Uranian ring system includes three broad rings ($\zeta$, $\nu$, and $\mu$) and ten narrow rings (6, 5, 4, $\alpha$, $\beta$, $\eta$, $\gamma$, $\delta$, $\lambda$, and $\epsilon$). The innermost Uranian moons, Cordelia and Ophelia, each with diameters of $\sim 40$ km \citep{Kark01}, flank the $\lambda$ and $\epsilon$ rings and play a role in shepherding the $\epsilon$ ring \citep{French91}. 
The narrow rings except the $\lambda$ ring are optically thick at visible wavelengths and are expected to be dominated by centimeter- to meter-sized particles \citep{Nicholson18}. Several small moons of diameters 40-135 km (Ophelia, Bianca, Cressida, Desdemona, Juliet, and Portia; \citealt{Kark01}) are located between the $\epsilon$ and $\nu$ rings, and several more, of diameters 20-160 km (Rosalind, Cupid, Belinda, Perdita, Puck, and Mab \citealt{Kark01, Showalter06}), between the $\nu$ and $\mu$ rings. Portia and Puck are the largest moons in these regions, with diameters of 135 and 160 km, respectively. Beyond Mab are the five largest Uranian moons: Miranda, in a class of its own with a diameter of 470 km; and then Ariel, Umbriel, Titania, and Oberon, with diameters of 1150-1580 km \citep{Thomas88}. 

\begin{table}
\begin{tabular}{lrr p{9cm}}
\hline
\toprule
moon & $a$ (km) & $e$ \\
\midrule
\textbf{moons of Uranus} \\
Cordelia  & 49,752 & 0.00026\\
Ophelia   & 53,763 & 0.00992\\
Bianca    & 59,166 & 0.00092\\
Cressida  & 61,767 & 0.00036\\
Desdemona & 62,658 & 0.00013\\
Juliet    & 64,358 & 0.00066\\
Portia    & 66,097 & 0.00005\\
Rosalind  & 69,927 & 0.00011\\ 
Cupid     & 74,392 & -\\
Belinda   & 75,256 & 0.00007\\
Perdita   & 76,417 & 0.00329\\
Puck      & 86,004 & 0.00012\\
Mab       & 97,736 & 0.00254\\
\midrule
\textbf{moons of Neptune} \\
Naiad     & 48,228 & 0.00014\\
Thalassa  & 50,075 & 0.00019\\
Despina   & 52,526 & 0.00027\\
Galatea   & 61,953 & 0.00020\\
Larissa   & 73,548 & 0.00121\\
Hippocamp &105,253 & 0.00001\\
Proteus   &117,647 & 0.00047\\
\hline
\end{tabular}
\caption{Semi-major axes $a$ and eccentricities $e$ of the inner moons of Uranus, 
from \citet{Jacobson98} and \citet{Showalter06}; and of Neptune, from \citet{Brozovic20}. 
\label{inner_moons} }
\end{table}
\begin{table}

\begin{tabular}{crr p{9cm}}
\hline
ring & $\bar{a}$ (km) & $\bar{a}e$ (km) \\
\midrule
\textbf{narrow rings of Uranus} \\
6 &41,838 & 43\\
5 &42,235 & 80 \\
4 &42,572 & 45\\
$\alpha$ &44,719 & 34\\
$\beta$ &45,661 & 20\\
$\eta$ &47,176 & -\\
$\gamma$ &47,627 & 5\\
$\delta$ &48,301 & -\\
$\lambda$ &50,024 & -\\
$\epsilon$ &51,150 & 406\\
\textbf{broad rings of Uranus} & &$W$ (km)  \\
$\zeta$ (Voyager) &38,300 &2500 \\ 
$\zeta$ (Keck) &39,600 &3500 \\
$\nu$ &67,300 & 3800\\
$\mu$ &97,700 &17,000\\
\midrule
\textbf{rings of Neptune} & &$W$ (km)  \\
Galle      &42,000 & 2,000 \\
Le Verrier &53,200 & 100 \\
Lassell    &55,200 & 4,000 \\
Arago      &57,200 & - \\
Galatea co-orbital &61,953 & - \\
Adams & 62,933 & 15 (in arcs) \\
\hline
\end{tabular}
\caption{Parameters of the rings of Uranus, from \citet{Nicholson18}; and of Neptune, from \citet{DePater18}. $\bar{a}$ is the mean semi-major axis. 
\label{rings} }
\end{table}

With such a quantity of moons exterior to the narrow rings, it is important to distinguish planetary normal mode resonances from resonances with satellites. 
A wave generated by a Lindblad resonance with an exterior moon propagates outward, whereas a wave generated by a Lindblad resonance with an interior moon or with most planetary normal modes propagates inward. 
Because we know where the moons are, we can calculate where their Lindblad resonances fall. 
Some of these line up nicely with some of the outer rings of Uranus, but none of them line up well with the inner rings \citep{Hedman21}. 
For example, the 6:5 resonance with Ophelia falls within the $\gamma$ ring \citep{Porco87, Hedman21}, even though its kinematics also includes $m=0$ and $m=1$ normal modes \citep{French91}.
The $\epsilon$ ring is shepherded by Cordelia and Ophelia: its inner edge coincides with the 24:25 outer eccentric resonance with Cordelia, while the outer edge coincides with the 14:13 inner eccentric resonance with Ophelia \citep{Goldreich87, French91, French95, Nicholson18}. 

The $\eta$ ring's kinematics are influenced by a 3:2 inner Lindblad resonance with Cressida \citep{Chancia17}. 
Finally, the $\delta$ ring's kinematics are well modeled by a single $m=2$ normal mode of the ring itself \citep{French91}.
Nevertheless, no known satellite resonances are found close to the 6, 5, 4, $\alpha$, and $\beta$ rings.

Neptune has one large moon, Triton, which orbits with a high inclination and in the retrograde direction. 
Proteus orbits Neptune about one third the distance to Triton and is less than 1/500 the mass of Triton \citep{Davies91, Stooke94}, but is Neptune's next largest moon and can be considered the outermost of the inner moons.
The other inner moons, from outward in, are the recently discovered 35-km diameter moon Hippocamp \citep{Showalter19}; then three moons of diameters of 150-200 km, namely, Larissa, Galatea, and Despina; and finally Thalassa and Naiad, which have diameters of 60-80 km \citep{Kark03}. 

\citet{DePater18} recently reviewed the current state of our knowledge of Neptune's rings.
Neptune has two broad faint rings, the Galle ring and the Lassell ring, 
and four narrow rings: the Le Verrier ring, the Arago ring, an unnamed ring that is co-orbital with Galatea, and the Adams ring (see Table \ref{rings}). 
The optical depth of the Le Verrier ring is comparable to that of the Adams ring outside of the arcs, while the optical depth of the Galle and Lassell rings is two orders of magnitude lower \citep{Porco95}.
It is still not entirely clear how any of these narrow rings are confined, and the rings seem to lack the fine-scale structure that Saturn's rings have.


\section{Methods \label{meth}}

In this Section, we first describe the interior models we use in Subsection \ref{models}. 
Next we explain how our frequency calculations account for rotation in Subsection \ref{freq}.
Then we show how we identify the sources of gravitational potential perturbations in Subsection \ref{gravpot}.
Last, we describe how we calculate the resonance locations in Subsection \ref{rescalc}.

\subsection{Interior Models \label{models}}

\begin{figure*}[htb] 
   \centering
   \includegraphics[width=\textwidth, trim={1.75in 1.8in 0 0.3in}]{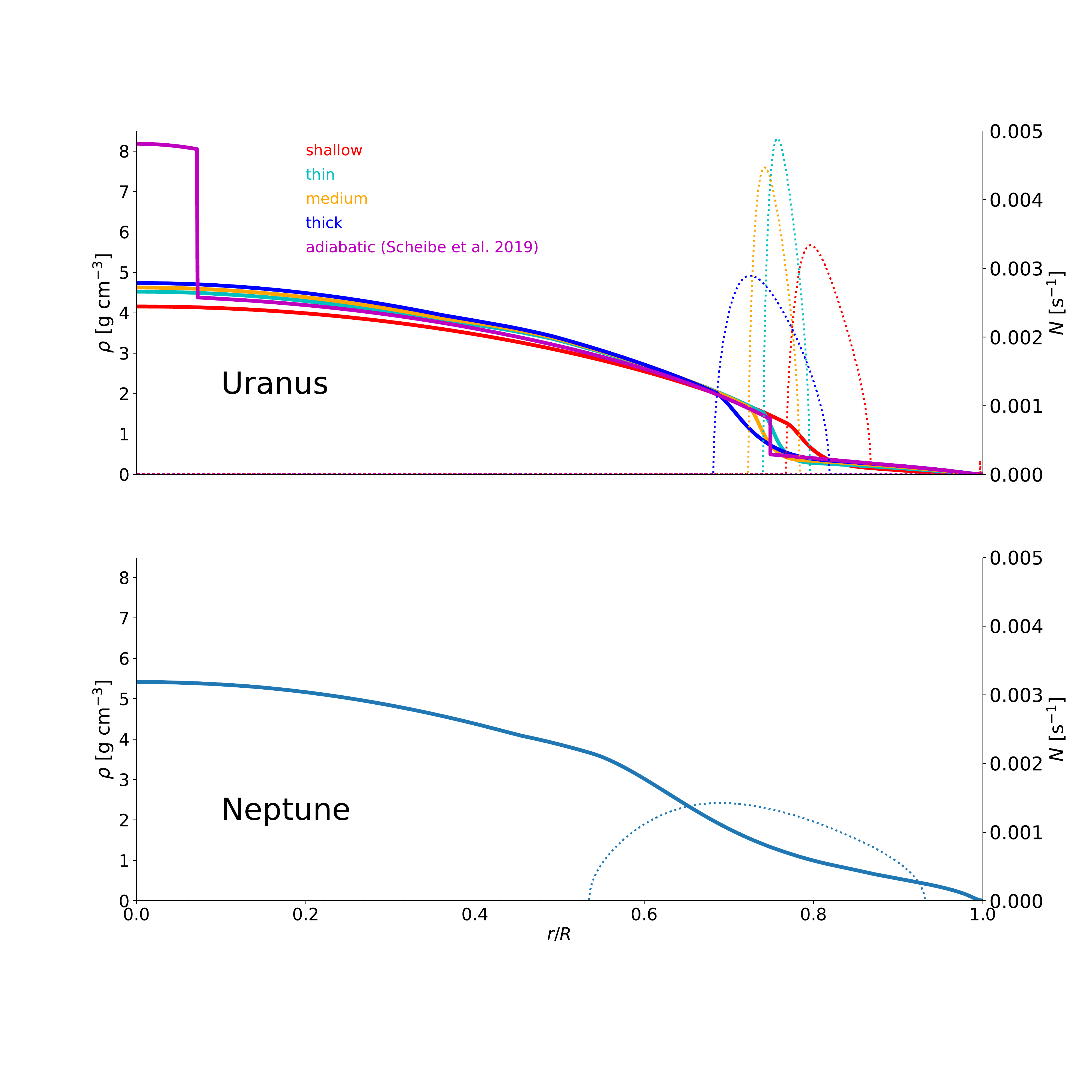} 
   \caption{Density (solid lines) and Brunt-V\"ais\"al\"a frequency (dotted) for models of Uranus (top) and Neptune (bottom) as functions of the fractional radius $r/R$. The scales of the $y$-axes are set equal for both the top and bottom panels for easier comparison between Uranus and Neptune models. 
   }
   \label{density_and_BV}
\end{figure*}

The adiabatic normal mode spectrum for a star or planet depends on the profile of mass density $\rho$, adiabatic sound speed $c$, and rotation rate $\Omega$.
The adiabatic sound speed $c$ is defined \citep{Unno79}

\begin{equation} 
    c^2 = \Gamma_1 \frac{P_0}{\rho_0},
\end{equation}
where $P_0$ and $\rho_0$ are the pressure and density of the unperturbed state, and the adiabatic exponent $\Gamma_1$ is

\begin{equation}
    \label{Gamma1}
    \Gamma_1 = \left(\frac{\partial \ln P}{\partial \ln \rho} \right)_{\rm{ad}}. 
\end{equation} 
Non-adiabatic regions are characterized by a non-zero Brunt-V\"ais\"al\"a frequency $N$: 

\begin{equation}
N^2 = -\frac{GM_r}{r}\left( \frac{d \ln \rho}{d \ln r} - \frac{1}{\Gamma_1}\frac{d \ln P}{d\ln r} \right),
\end{equation}
where $G$ is the gravitational constant, $M_r$ is the mass interior to the radius $r$. 
For clarity, $M$ without the subscript $r$ will refer to the total mass of the planet.
The Brunt-V\"ais\"al\"a frequency $N$ quantifies the angular frequency with which a small parcel of gas oscillates radially with positive or negative buoyancy under local pressure balance with its surrounding gas \citep{Unno79}.

Here we consider spherically symmetric models so that $\rho=\rho(r)$, $c=c(r)$, and make the further assumption of rigid rotation so that $\Omega(r)=\Omega={\rm constant}$.
Although the cloud-level jet streams alone guarantee some degree of differential rotation in Uranus and Neptune \citep{Kaspi13}, the uncertainty in their underlying bulk rotation rates means that rigidly rotating models are sufficient for our purposes.

The system of equations required to solve for the frequencies of the normal mode spectrum is found in classic works like \citet{Unno79},
and there is now  a publicly available and extensively validated asteroseismology software package called {\tt GYRE}
that solves these equations \citep{Townsend13}.  
This code has recently been used successfully for Saturn \citep{Mankovich19, Markham20}.

Most models of the interiors of the ice giants make a number of simplifying assumptions, such as a three-layer structure: 
a rock-rich core, a water-rich envelope, and a hydrogen-rich atmosphere. 
Many models also assume an adiabatic interior (see, for example, \citealt{Scheibe19}).
Thermal evolution models, however, suggest that the classical assumption of an adiabatic interior is inconsistent with the luminosities of Uranus and Neptune, and for this reason, models are being explored that are not fully adiabatic, but instead have a thermal boundary layer (see, for example, \citealt{Scheibe21, Stixrude21}).
The gravitational harmonics determined by the Voyager flybys and observations of the dynamics of rings and moons as well as theoretical considerations based on laboratory experiments work together to constrain the properties of the planets' layers. 
Some fundamental aspects of ice giant interiors, such as the ice-to-rock mass fraction, are poorly constrained \citep{Helled20, Podolak19}.
Even the rotation rates measured by Voyager for Uranus \citep{Desch86, Warwick86} and Neptune \citep{Warwick89} have been called into question \citep{Helled10}.
Given these uncertainties, it is worth considering a relatively broad range of models.

Figure \ref{density_and_BV} shows profiles of density and Brunt-V\"ais\"al\"a frequency of Uranus and Neptune interior models that we use in this work. 
What we have labeled the adiabatic model is from \citet{Scheibe19}.
Adiabatic oscillation calculations in {\tt GYRE} require the thermodynamic derivatives $\Gamma_1$ (Eq. \ref{Gamma1}) and 

\begin{equation}
    \nabla_{\rm ad} = \frac{1}{\Gamma_1}\left( \frac{\partial \textrm{ln} T}{\partial \textrm{ln} \rho} \right)_{\rm ad},
\end{equation}
which we calculate numerically from the adiabatic model's $\rho(P)$ and $T(P)$. 
Because this model assumes a three-layer structure and features density discontinuities owing to sudden composition changes, double mesh points were inserted at the core-envelope and inner-outer envelope boundaries ($r/R\sim 0.07$ and $r/R \sim 0.75$, respectively, where $R$ is the planetary radius) in order for {\tt GYRE} to apply the appropriate jump conditions at those locations.

We considered a Saturn model from \citet{Mankovich21} and confirmed that our mode calculation reproduced results consistent with theirs, up to the expected error associated with our first-order treatment of rotation (see below).

We have also constructed new models of Uranus and Neptune using the method described in \citet{Mankovich21}. 
In brief, these models are calculated using the fourth-order theory of figures \citep{Nettelmann2017}, assuming H-He-H$_2$O mixtures modeled using the MH13-SCvH equation of state for hydrogen and helium \citep{Militzer2013,Saumon1995,Miguel2016} and \citet{Mazevet19} for water, combined in an additive-volume approximation. The \citet{Mazevet19} equation of state for water transitions to that of an ideal gas below $T=800$ K, following \citet{Scheibe19}.
Rather than the common practice of imposing discontinuous changes in composition, these models impose a gradient in the water abundance between a homogeneous H-He dominated outer envelope and a homogeneous H$_2$O-dominated interior. 
Denoting the H, He and H$_2$O mass fractions by $X$, $Y$, and $Z$ so that $X+Y+Z=1$, we fix $Y/(X+Y)=0.275$ throughout the model following the protosolar value estimated by \citet{Asplund09}, and for $Z(r)$ assume a sigmoid function with four free parameters specifying the inner and outer radii of the gradient region and the $Z$ values at those boundaries (see \citealt{Mankovich21} for details). One of these parameters is eliminated by the condition that the model satisfy each planet's equatorial radius. 

Despite the stabilizing influence of the composition gradient, these models assume for simplicity that temperature is adiabatically stratified, subject to the boundary condition that $T=150$ K at $P=10$ bar, consistent with atmosphere models \citep{Fortney2011}. 

We present mode calculations for three Uranus models, one fitting $J_2$ and $J_4$ exactly, and two that are offset in $J_4$ relative to the observed value by approximately $\pm1$ times the measurement uncertainty $\sigma_{J_4}=1.30 \times 10^{-6}$ \citep{Jacobson14}. The models achieve this by varying the width of the $Z$ gradient region and are accordingly labeled ``thin'', ``medium'' and ``thick'', corresponding to $J_4$ offsets of $-0.89$, 0, and $+0.79$ times $\sigma_{J_4}$, respectively.
Another Uranus model we label ``shallow'' has a $Z$ gradient region closer to the planet's surface.
For Neptune we present a single model that fits $J_2$ and $J_4$. These models are summarized in Table~\ref{model_params} and their interior structures shown in Figure~\ref{density_and_BV}, where the stably stratified composition gradient regions are visible as regions with positive Brunt-V\"ais\"al\"a frequency. 
Although we are primarily focusing our discussion on f-modes, the models with stably stratified composition regions also generate g-modes, which in principle can mix with the f-modes, and can also be associated with their own resonances.


\subsection{Normal Mode Frequency Calculation \label{freq}}

Because the ratios of rotation rate to breakup frequency of Uranus and Neptune, 
\begin{equation}
    \frac{\Omega}{\sqrt{\frac{GM}{R^3}}} = \begin{cases}
      0.175 \pm 0.004, & \text{Uranus} \\
      0.155 \pm 0.006, & \text{Neptune} 
    \end{cases} 
\end{equation}
(with uncertainties dominated by uncertainties in the bulk rotation rate; see \citealt{Desch86, Warwick86, Warwick89, Helled10}), 
are significant, Coriolis accelerations are not negligible compared to the other terms in the momentum equation. 
For this reason, the oscillation frequencies given by {\tt GYRE} are corrected to first order in the planet's rotation rate $\Omega$ \citep{Ledoux51, Unno79}.
These corrections account for the Doppler shift and the approximate intrinsic perturbation to mode frequencies due to the Coriolis force.

{\tt GYRE} evaluates an integral to provide the rotation splitting coefficient $\beta$ that is involved in the Coriolis perturbation for solid-body rotation. 
$\beta$ can also be calculated for a given mode with the equation (cf. \citealt{Unno79}) 
\begin{equation}
    \beta = 1 - \frac{\int_0^R \left(2 \xi_r \xi_h + \xi_h^2 \right) \rho r^2 dr }{\int_0^R \left[ \xi_r^2 + \ell \left( \ell + 1 \right) \xi_h^2 \right] \rho r^2 dr}
\label{eq_beta}
\end{equation}
where $\xi_r$ and $\xi_h$ are the radial and horizontal components of the displacement eigenfunction \citep{Unno79}, which is given by

\begin{equation}
\begin{aligned}
    \mathbf{\xi} \left(r,\theta,\phi,t\right) = &\Biggl[ \xi_r \left(r\right) \hat{r} + \xi_h \left(r\right) \left( \hat{\theta} \frac{\partial}{\partial \theta} + \hat{\phi} \frac{1}{\sin \theta} \frac{\partial}{\partial \phi} \right) \Biggr] \\ &\times Y_{\ell}^m \left(\theta,\phi \right) e^{i \sigma_{\ell m n} t},
\end{aligned}
\end{equation}
We did verify that the $\beta$ values returned by {\tt GYRE} are consistent with this formula.

Because Uranus and Neptune rotate more slowly than Jupiter and Saturn, we do not include second-order terms associated with oblateness and the centrifugal force, which together tend to decrease the frequencies \citep{Vorontsov81}. For this reason, our uncertainties in calculated mode frequencies, and thus resonance locations, are one-sided.  

We calculate the moment of inertia for each model, assuming spherical symmetry, by integrating
\begin{equation}
    \label{inertia}
    \frac{I}{MR^2} = \frac{8 \pi}{3} \frac{\int \rho \left(r\right) r^4 dr}{MR^2}
\end{equation}
Table \ref{model_params} shows the parameters that were used for each model, as well as each model's calculated moment of inertia. 

\subsection{Sources of Gravitational Potential Perturbations \label{gravpot}}
Following \citet{Marley93}, we can write the total gravitational potential as a sum of an unperturbed component and a perturbed component:

\begin{equation}
    \Phi = \Phi_0 + \Phi \sp{\prime} \left( t \right),
\end{equation}
where full expressions for the unperturbed component $\Phi_0$ and for the perturbed component $\Phi \sp{\prime} \left( t \right)$ can be found in \citet{Marley93}.
The integrals for the perturbed gravitational harmonics are taken over the Eulerian density perturbation

\begin{equation}
\rho_{\ell m n} \sp{\prime} = \rho_{\ell n} \sp{\prime} \left( r \right) Y_{\ell}^{m} \left(\theta,\phi \right) e^{-i \sigma_{\ell m n} t}
\end{equation}
instead of over the unperturbed density $\rho$.
In the above equation, $\sigma_{\ell m n}$ is the oscillation frequency in the reference frame that rotates with the planet, $r$ is the radius, $\theta$ is the colatitude, and $\phi$ is the azimuthal angle. 
The spherical harmonics $Y_{\ell}^m$ are defined in terms of the associated Legendre polynomials $P_{\ell}^m$. 

The equations for the gravitational harmonics that appear in the equations for $\Phi \sp{\prime} \left( t \right)$ can be reduced to 

\begin{equation}
    M R^{\ell} J_{\ell n} \sp{\prime} = - \left( \frac{4 \pi}{2 \ell + 1} \right)^{1/2} e^{i \sigma_{\ell 0 n} t} \int_0^{R} \rho_{\ell n} \sp{\prime} \left(r\right) r^{\ell + 2} dr
\label{J_eq}
\end{equation}
for $m=0$, and 

\begin{equation}
\begin{aligned}
    M R^{\ell} C_{\ell m n} \sp{\prime} = &\left( -1 \right)^{\frac{m+|m|}{2}} \Biggl[ \frac{2 \ell + 1}{4 \pi} \left( \frac{ \left( \ell - |m| \right)!}{\left( \ell + |m| \right)!} \right) \Biggr]^{1/2} \\ &\times e^{i \sigma_{\ell m n} t} \int_0^{R} \rho_{\ell m n} \sp{\prime} \left(r\right) r^{\ell + 2} dr
\end{aligned}
\label{C_eq}
\end{equation}
for $m \ne 0$.
$S_{\ell m n} \sp{\prime}$ can be made to vanish by an appropriate choice of phase \citep{Marley93}. 

\begin{figure}[htb] 
   \centering
   \includegraphics[width=\columnwidth, trim={0.75in 0 0.75in 0.25in}]{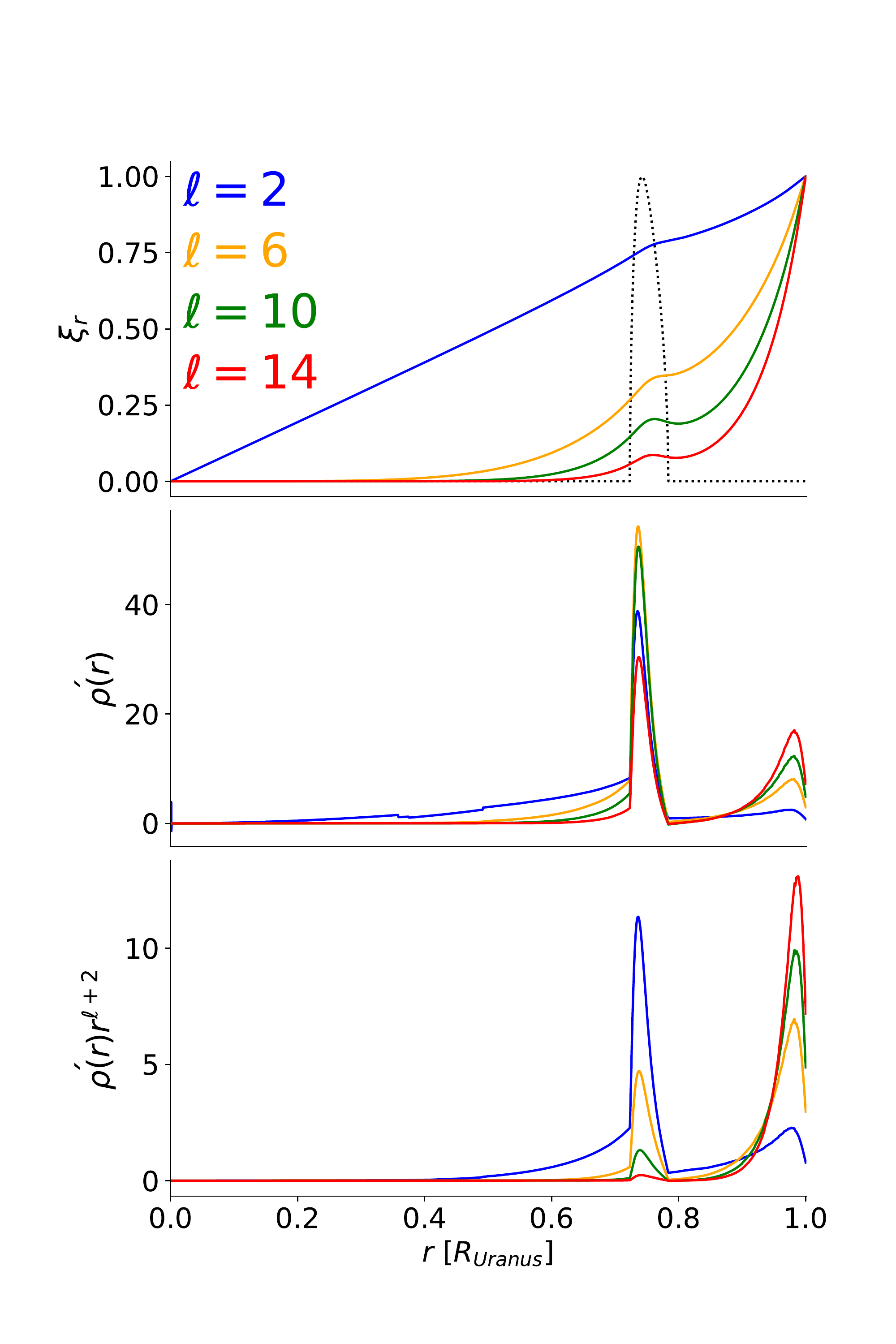} 
   \caption{(Top) Radial displacement eigenfunctions for $\ell=2,6,10$, and $14$, normalized to $\xi_r = 1$ at $r = R$, from the medium Uranus model. The Brunt-V\"ais\"al\"a frequency $N$, normalized to peak at 1, is shown with dotted lines. (Middle) Eulerian density perturbation $\rho \sp{\prime} \left(r\right)$ for the same four modes. (Bottom) Integrand of Equations \ref{J_eq} and \ref{C_eq}, $\rho \sp{\prime} r^{\ell + 2}$, relevant for gravitational potential perturbations, for the same modes. The modes of lower oscillation degree $\ell$ are more sensitive to inner layers of the planet, while the modes of higher oscillation degree $\ell$ are more sensitive to outer layers of the planet.
   }
   \label{xi_plot}
\end{figure}

The radial displacement eigenfunctions of $\ell=2,6,10,$ and $14$ from our medium model are plotted in the top panel of Figure \ref{xi_plot}, normalized to $\xi_r = 1$ at $r = R$.
We also show the normalized Brunt-V\"ais\"al\"a frequency $N$ as a dotted line to highlight its correlation with the warp in the radial displacement eigenfunctions. 
The middle panel of Figure \ref{xi_plot} shows the Eulerian density perturbation of the same modes $\ell=2,6,10,$ and $14$. 
The bottom panel of Figure \ref{xi_plot} shows the integrand of Equations \ref{J_eq} and \ref{C_eq}, $\rho \sp{\prime} r^{\ell + 2}$, for each of these modes. 
These plots illustrate how the modes of lower spherical harmonic degree $\ell$ can probe deeper into the planet, whereas the modes of higher spherical harmonic degree $\ell$ are more sensitive to the outer layers and to the parts of the planet that are stably stratified. 

Only the spherical harmonic $Y_{\ell}^0$ contributes to the $J_{\ell n} \sp{\prime}$ term due to the orthogonality of the spherical harmonics. 
Likewise, for $m \ne 0$, the contribution to each $C_{\ell m  n} \sp{\prime}$ term is only from the $Y_{\ell}^m$ harmonic.
Each oscillation mode thus contributes to a single perturbed gravitational harmonic. 
The planet's rotation, however, can mix modes slightly. 
For Saturn, no more than 15\% of the rotationally corrected radial displacement eigenfunction could be attributed to mode mixing \citep{Marley93}; for Uranus and Neptune, which rotate more slowly than Saturn, we can expect mode mixing to influence the radial displacement eigenfunction to an even lesser extent. 

\subsection{Resonance Calculation \label{rescalc}}

We calculate the pattern frequency $\Omega_{\rm{pat}}$ of each normal mode seen from inertial space, 

\begin{equation}
    \Omega_{\rm{pat}} = \frac{1}{m} \left( \sigma_{\ell n}^0 + m \beta \Omega \right),
\end{equation}
where $\sigma_{\ell n}^0$ is the oscillation frequency in the non-rotating limit for the mode specified by $\ell$ and $n$, and $\beta$ is the rotation splitting coefficient defined in Equation \ref{eq_beta}.
Our models for Uranus assume the rotation rate provided in \citet{Helled10}, called the fast rotation rate, corresponding to a period of 16.58 hours. The adiabatic model assumes the classical, or slow, rotation rate, corresponding to a period of 17.24 hours \citep{Desch86, Warwick86}. Our Neptune model likewise uses the rotation rate from \citet{Helled10}, corresponding to a period of 17.46 hours, which in contrast is slower than the classical rotation rate provided by Voyager 2 radio data, corresponding to a period of 16.11 hours \citep{Warwick89}. In general, resonance locations of a faster rotator fall further inward than those of a slower rotator.  
Thus the rotation rate of these planets is another parameter that ring seismology could help constrain, similar to how \citet{Mankovich19} calculated a seismological rotation rate for Saturn.
\begin{table*}
\begin{tabular}{lrrrrrrr p{9cm}}
\hline
model & $R$ (km) & $M$ ($\times 10^{25}$ kg) & $\Omega$ ($\times 10^{-4}$ s$^{-1}$) & $J_2$ ($\times 10^{-4}$) & $J_4$ ($\times 10^{-4}$) & $I/MR^2$ \\
\hline
Uranus shallow  & 25,559 &  8.68009 & 1.0527 & 35.1068 & -0.341705 & 0.2279 \\
Uranus thin  & 25,559 &  8.68009 & 1.0527 & 35.1068 & -0.330171 & 0.2204 \\
Uranus medium  & 25,559 &  8.68009 & 1.0527 & 35.1068 & -0.341705 & 0.2202 \\
Uranus thick  & 25,559 &  8.68009 & 1.0527 & 35.1069 & -0.351949 & 0.2200 \\
Uranus adiabatic$^{\dagger}$  & 25,559 &  8.68009 & 1.0124 & 35.107 & -0.342 & 0.2266 \\
Neptune & 24,764 & 10.24092 & 0.9996 & 34.0843 & -0.334 & 0.2509 \\
\hline
\end{tabular}
\caption{Parameters of the planetary models. 
Equatorial radii $R$ are from \citet{Archinal18}.
$M$, $J_2$, and $J_4$ of Uranus are from \citet{Jacobson14}, 
while $M$, $J_2$, and $J_4$ of Neptune are from \citet{Jacobson09}. 
The classical Uranus spin rate $\Omega$ is from Voyager 2 radio data \citep{Desch86, Warwick86} and is used only in the adiabatic model, while alternate spin rates $\Omega$ based on the shapes and gravitational coefficients of the planets are from \citet{Helled10} and are used in the other models.
The classical Neptune spin rate $\Omega$, not used in any of our models, is $1.0834 \times 10^{-4}$ s$^{-1}$ \citep{Warwick89}.
The moments of inertia are calculated using Equation \ref{inertia}.
\\ $^{\dagger}$ \citet{Scheibe19}
\label{model_params} }
\end{table*}

Then we calculate the resonance location numerically. 
Lindblad resonances occur at locations where the following relationship is satisfied \citep{Goldreich79}: 

\begin{equation}
    m \left( n - \Omega_{\rm{pat}}\right) = \pm q \kappa,
\end{equation}
where the upper sign corresponds to an inner Lindblad resonance and the lower sign corresponds to an outer Lindblad resonance, $q$ is a positive integer, and $n$ and $\kappa$ are the resonance location's mean motion and horizontal epicyclic frequency, respectively. These are calculated according to the second-order equations from \citet{Renner06}.


Similarly, vertical resonances occur at locations where 

\begin{equation}
    m \left( n - \Omega_{\rm{pat}}\right) = \pm b \mu,
\end{equation}
where the upper sign corresponds to an inner vertical resonance and the lower sign corresponds to an outer vertical resonance, $b$ is a positive integer, and $\mu$ is the resonance location's vertical epicyclic frequency, which is found using the equation from \citet{Shu83}:

\begin{equation}
    \mu^2 + \kappa^2 = 2n^2
\end{equation}
We are focusing only on first-order resonances, which have $q=b=1$, though higher-order resonances are possible \citep{Marley14}. 

Corotation resonances are potentially interesting to consider in the context of Neptune's Adams ring (see Section \ref{nep_lev_adams}). Hence we compute corotation resonance locations for Neptune following the same procedures as in \citet{A21}. That is, we find the radii where the mean motion $n$ matches the pattern frequencies $\Omega_{\rm{pat}}$ associated with the $\ell=m$ modes we have calculated from our Neptune model.

\section{Results \label{res}}

\begin{figure*}[htb] 
   \centering
   \includegraphics[width=\textwidth, trim={1.0in 0 1.0in 0}]{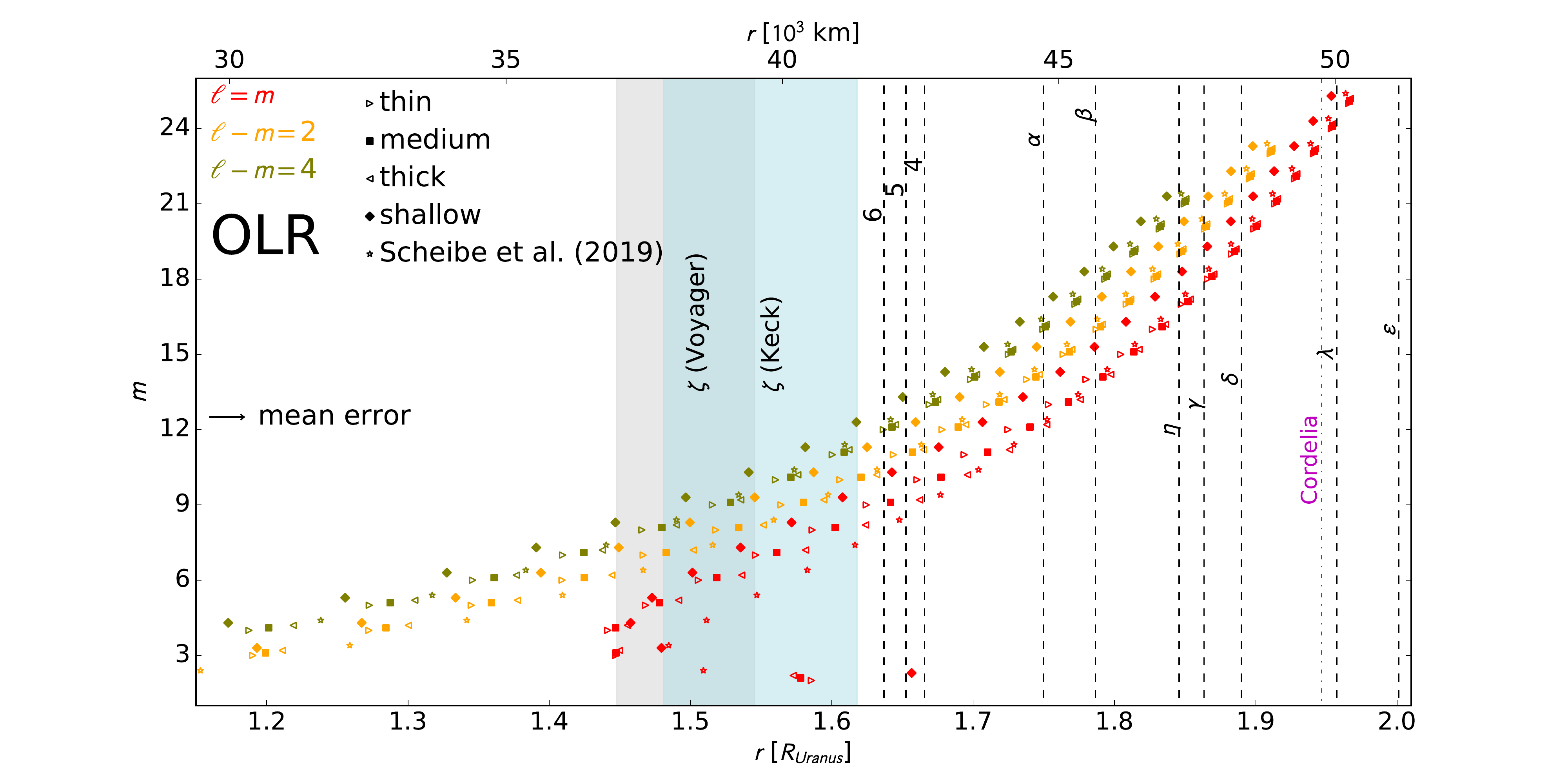} 
   \caption{Normal mode outer Lindblad resonance (OLR) location predictions from all five Uranus models. The azimuthal order $m$ (equivalent to the number of spiral arms) is the vertical axis, and distance from the center of Uranus is the horizontal axis, shown in Uranus radii on the bottom and in km on the top. OLRs can excite inward-propagating spiral density waves. The typical uncertainty of these resonance locations is shown by one bar on the left.
   }
   \label{U_OLR}
\end{figure*}

\begin{figure*}[htb] 
   \centering
   \includegraphics[width=\textwidth, trim={1.0in 0 1.0in 0}]{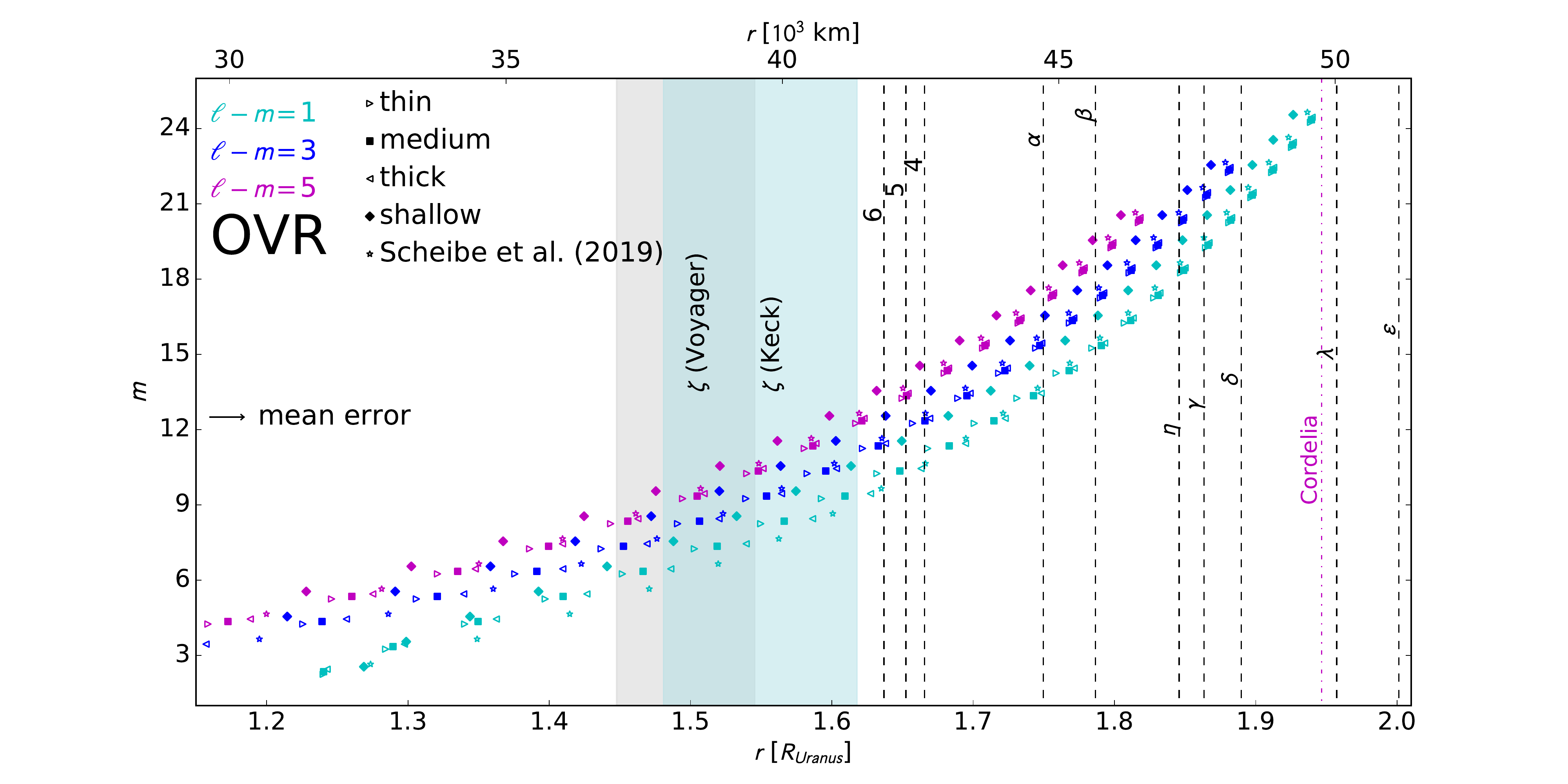} 
   \caption{Normal mode outer vertical resonance (OVR) location predictions from all five Uranus models. The azimuthal order $m$ (equivalent to the number of spiral arms) is the vertical axis, and distance from the center of Uranus is the horizontal axis, shown in Uranus radii on the bottom and in km on the top. OVRs can excite outward-propagating bending waves. The typical uncertainty of these resonance locations is shown by one bar on the left.
   }
   \label{U_OVR}
\end{figure*}

Figures \ref{U_OLR} and \ref{U_OVR} show the calculated resonance locations with f-mode oscillations for all our Uranus models. Figure \ref{U_OLR} shows the Lindblad resonances, while Figure \ref{U_OVR} shows the vertical resonances.
We show the resonance locations up to $\ell=25$.
For every mode, the resonance location of the medium model falls in between those of the thin and thick models. 
The resonance location of the thick model typically falls further outward (and the resonance location of the thin model falls further inward), the $\ell=m=2$ mode being the sole exception.  
This general trend is set by the models' different predictions for density in the outer envelope, which the higher $m$ modes are more sensitive to (cf. Figure \ref{xi_plot}). 

The region of particular interest that we find spans from the $6$ ring to the $\beta$ ring. 
Satellite resonances do not line up with these inner rings \citep{Hedman21},
but several high-$m$ mode resonances from our models do fall in this region. 
The predictions from our models as well as from \citet{Marley88} include resonance locations that fall in the $\zeta$ ring for the modes with $\ell \leq 5$.
The $\zeta$ ring, however, has low optical depth, and would not be expected to sustain a wave. 

Lindblad resonances that fall among the inner rings are listed in Table \ref{L_inner_rings}, whereas vertical resonances are listed in Table \ref{v_inner_rings}.
The resonance locations are given in general ranges, based on all our models as well as the frequency uncertainty from our calculations with first-order rotation corrections. 
Predictions based on individual interior models are found in the Appendices. 
We group the 6, 5, and 4 rings together, and the $\alpha$ and $\beta$ rings together, in these tables, because the uncertainty associated with a mode's frequency or resonance location is comparable to the distance between these narrow rings. 
Differential rotation confined to the outer layers of the interior will have a greater effect on the higher-degree modes.  

Among the 6, 5, and 4 rings, we expect that a Lindblad resonance with $m=2$ or $7 \leq m \leq 13$, or a vertical resonance with $9 \leq m \leq 14$, could be from a planetary normal mode. Among the $\alpha$ and $\beta$ rings, we expect that a Lindblad resonance with $11 \leq m \leq 18$, or a vertical resonance with $13 \leq m \leq 19$, could be from a planetary normal mode. 

\begin{table}
\begin{tabular}{rrrr p{9cm}}
\toprule
$\ell$ &$m$ &$\Omega_{\rm{pat}}$ (deg/day) &$r_{\rm{res}}$ (km)  \\
\midrule
\multicolumn{2}{c}{6, 5, or 4 ring} & \multicolumn{2}{r}{41,838; 42,235; 42,572} \\
2  &2  &1995.2-2360.4 &38571-43142 \\
7  &7  &1585.4-1753.2 &39244-41963 \\
8  &8  &1517.2-1666.7 &40166-42760 \\
9  &9  &1460.4-1591.0 &41088-43498 \\
10 &10 &1412.3-1525.1 &41981-44184 \\
12 &10 &1506.0-1605.7 &40565-42334 \\
13 &11 &1450.1-1537.1 &41532-43175 \\
14 &12 &1402.8-1479.3 &42409-43935 \\
16 &12 &1466.5-1537.2 &41338-42655 \\
17 &13 &1419.4-1482.8 &42176-43420 \\
\midrule
\multicolumn{2}{c}{$\alpha$ or $\beta$ ring} & \multicolumn{2}{r}{44,719; 45,661} \\
11 &11 &1370.9-1467.8 &42827-44822 \\
12 &12 &1334.6-1418.1 &43618-45417 \\
13 &13 &1301.1-1375.0 &44350-46012 \\
14 &14 &1271.6-1337.3 &45024-46562 \\
15 &15 &1245.4-1304.3 &45644-47070 \\
16 &14 &1326.8-1387.5 &43933-45262 \\
17 &15 &1295.6-1350.5 &44598-45850 \\
18 &16 &1267.7-1318.0 &45211-46397 \\
19 &15 &1342.5-1395.1 &43644-44776 \\
20 &16 &1310.5-1359.2 &44294-45382 \\
21 &17 &1281.9-1327.2 &44897-45949 \\
22 &18 &1256.1-1298.6 &45460-46480 \\
\hline
\end{tabular}
\caption{Predicted Lindblad resonance locations among the inner rings of Uranus. 
The ranges in frequencies and locations take into account the two most extreme models and include the error in the range given.  
\label{L_inner_rings} }
\end{table}

\begin{table}
\begin{tabular}{rrrr p{9cm}}
\toprule
$\ell$ &$m$ &$\Omega_{\rm{pat}}$ (deg/day) &$r_{\rm{res}}$ (km) \\
\midrule
\multicolumn{2}{c}{6, 5, or 4 ring} & \multicolumn{2}{r}{41,838; 42,235; 42,572} \\
10 &9  &1517.1-1642.0 &40245-42420 \\
11 &10 &1460.6-1566.9 &41241-43216 \\
12 &11 &1412.5-1503.4 &42159-43945 \\
14 &11 &1486.3-1569.6 &40966-42481 \\
15 &12 &1435.2-1508.5 &41868-43281 \\
17 &12 &1496.8-1565.4 &40850-42086 \\
18 &13 &1446.7-1508.4 &41704-42882 \\
19 &14 &1403.1-1459.4 &42486-43614 \\
\midrule
\multicolumn{2}{c}{$\alpha$ or $\beta$ ring} & \multicolumn{2}{r}{44,719; 45,661} \\
14 &13 &1332.1-1402.8 &43769-45302 \\
15 &14 &1299.6-1362.7 &44471-45896 \\
16 &15 &1270.9-1327.7 &45115-46446 \\
18 &15 &1319.4-1373.0 &44117-45303 \\
19 &16 &1289.4-1338.8 &44748-45881 \\
20 &17 &1262.6-1308.4 &45333-46422 \\
21 &16 &1331.0-1379.2 &43872-44922 \\
22 &17 &1300.7-1345.8 &44491-45511 \\
23 &18 &1273.4-1315.8 &45069-46063 \\
24 &19 &1248.7-1288.8 &45612-46582 \\
\hline
\end{tabular}
\caption{Predicted vertical resonance locations near the inner rings of Uranus. 
\label{v_inner_rings} }
\end{table}

Although planetary normal mode resonances with $m > 15$ may fall among the $\eta$, $\gamma$,  $\delta$, $\lambda$, and $\epsilon$ rings, we expect the dynamics of these rings to be influenced more by the moons Cordelia and Ophelia. 
The case has been made that while Cordelia is the outer shepherd of the $\delta$ ring and the inner shepherd of the $\epsilon$ ring, Ophelia is the outer shepherd of both the $\epsilon$ and $\gamma$ rings \citep{Porco87}.
Recently, the $\eta$ ring was found to be influenced by Cressida \citep{Chancia17}. 
Nevertheless, at Saturn the $\ell=m=10$ mode was found to be anomalously strong \citep{Hedman19}, and with this precedent the influence of planetary normal modes on the outer rings cannot be ruled out.

\begin{figure*}[htb] 
   \centering
   \includegraphics[width=\textwidth, trim={1.0in 0 1.0in 0}]{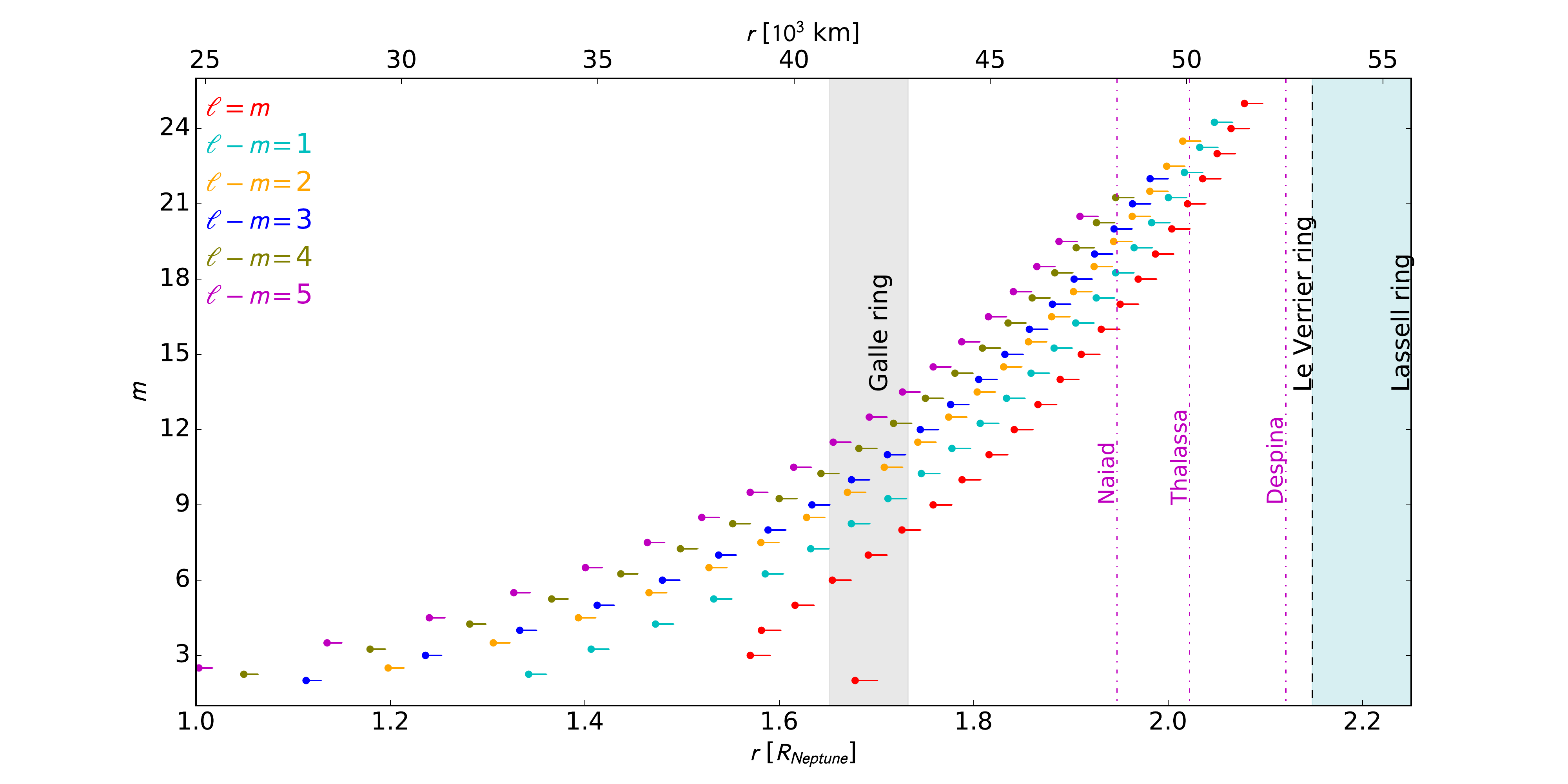} 
   \caption{Normal mode resonance location predictions for Neptune.
   The azimuthal order $m$ is the vertical axis, and distance from the center of Neptune is the horizontal axis, shown in Neptune radii on the bottom and in km on the top. Both  outer  Lindblad  resonances (OLR) and outer vertical resonances (OVR) are shown.
   Several resonances fall in the Galle ring, and others may perturb the inner moons Naiad and Thalassa.
   }
   \label{nep_predictions}
\end{figure*}

At Neptune, modes with $\ell=m=16,...,21$ fall among the moons Naiad and Thalassa. Naiad and Thalassa are themselves in a 73:69 resonance \citep{Brozovic20}, made possible by Naiad's high inclination (4.7$^{\circ}$). 
Should it be found that this resonance does not fully account for their orbits, perturbations from planetary normal modes may be part of the solution. 
Furthermore, it is not known how Naiad reached its high inclination. 
The most likely explanation is passage through a previous resonance with Despina \citep{Banfield92}.
Another possibility, however, is that Naiad's inclination was excited instead by a vertical resonance with a Neptunian normal mode. 

\begin{figure*}[htb] 
   \centering
   \includegraphics[width=\textwidth, trim={1.0in 0 1.0in 0}]{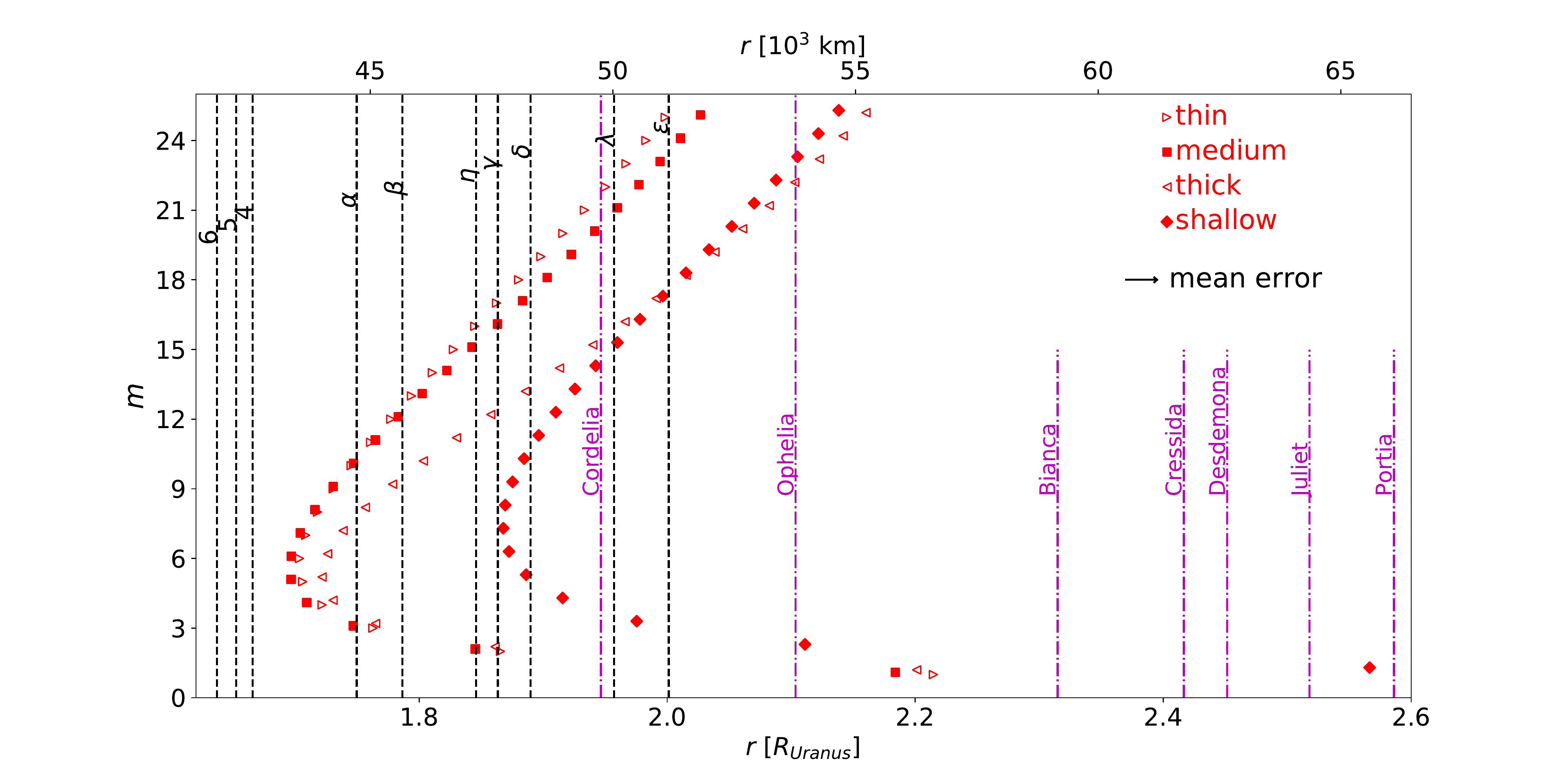} 
   \caption{Resonance location predictions for Uranian g-modes.
   The azimuthal order $m$ is the vertical axis, and distance from the center of Uranus is the horizontal axis, shown in Uranus radii on the bottom and in km on the top. Only  outer  Lindblad  resonances (OLR) with $\ell=m$, $n=1$ are shown.
   }
   \label{ura_predictions_g}
\end{figure*}

The above calculations focused exclusively on the f-modes  ($n=0$).
It is also interesting, however, to consider the $\ell=m$, $n=1$ g-modes for the non-adiabatic models because these can also fall among the ring systems.
In the Uranian system (Figure \ref{ura_predictions_g}), $\ell=m$, $n=1$ g-mode resonances are likely to fall in the mid- to outer-ring system as well as the innermost moons. 
The spread in resonance location predictions is greater for g-modes than for f-modes because the g-mode spectrum is sensitive mainly to the Brunt-V\"ais\"al\"a frequency $N(r)$. The thin-medium-thick triplet of models diverges more for higher $\ell$, where the eigenfunctions have higher amplitudes in the g-mode cavity.

\begin{figure*}[htb] 
   \centering
   \includegraphics[width=\textwidth, trim={1.0in 0 1.0in 0}]{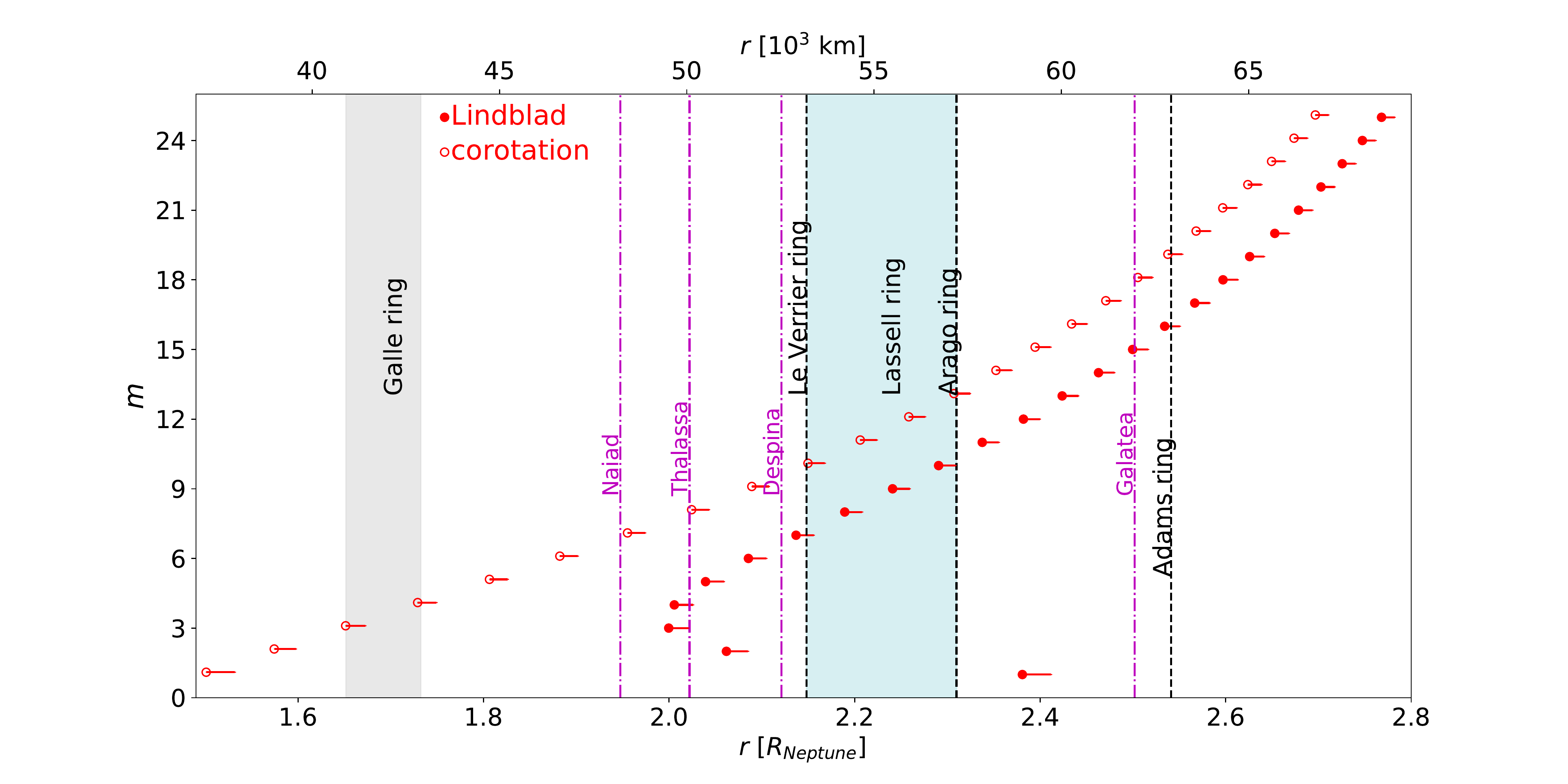} 
   \caption{Resonance location predictions for Neptunian g-modes.
   The azimuthal order $m$ is the vertical axis, and distance from the center of Neptune is the horizontal axis, shown in Neptune radii on the bottom and in km on the top. Outer  Lindblad  resonances (OLR) with $\ell=m$, $n=1$ are shown as filled-in circles, while corotation resonances associated with these Lindblad resonances are shown with open circles.
   }
   \label{nep_predictions_g}
\end{figure*}

In the Neptunian system (Figure \ref{nep_predictions_g}), $\ell=m$, $n=1$ g-mode resonances are likely to fall among the inner moons and middle to outer rings. 
There is even a g-mode resonance that falls near the Adams ring, which we discuss in Section \ref{nep_lev_adams}. 
G-modes at higher order ($n \geq 2$) are at lower frequency, and so their outer Lindblad resonance locations would fall farther out. Nevertheless, their gravity perturbations might be too small to matter because they are more effectively confined deep down than the $n=1$ g-modes.

Note that the mean absolute model uncertainty, due to the approximate treatment of rotation, in the resonance locations for all the f-modes we calculated is $\delta r = 610$ km for the Uranus models and $\delta r = 450$ km for the Neptune model.
The maximum uncertainty in f-mode resonance locations is  $\delta r = 805$ km for the Uranus models and $\delta r = 557$ km for the Neptune model. 
These both correspond to the $\ell=m=2$ f-mode. 
The uncertainty generally decreases with increasing $m$.

To decrease the uncertainty, frequency-correction calculations may be carried out to second-order using the perturbation theory techniques described in \citet{Vorontsov81} and summarized in the Appendix of \citet{Marley90}.
We leave the second-order calculations for possible future work. 
Further knowledge of the planets' interior structure, thermodynamic state, and rotational state can also improve the precision of resonance location predictions.

\section{Discussion \label{disc}}

We have found that planetary normal modes likely fall among the rings of Uranus and Neptune. The next question is whether normal mode resonances would be detectable in the Uranian and Neptunian rings.
Potential signatures of such resonances could include wave-like structures in the dense Uranian rings, correlations with Uranus's many narrow ring features, and longitudinal structures in more dusty rings.  Each of these methods is discussed in more detail below.

Note that we do not attempt to compute the explicit amplitudes of the gravitational perturbations associated with these perturbations.
While \citet{Marley93} estimated the strengths of the perturbations needed to produce the observed waves in Saturn's rings, more recent kronoseismology studies did not find strong correlations between the anticipated and the observed amplitudes of normal mode resonances \citep{Hedman19}. 
Although it is expected that torques decrease as the oscillation degree $\ell$ increases because a smaller quantity of mass participates in the oscillation \citep{Marley93},
the intrinsic amplitude spectrum of normal modes in the planet is unknown but the subject of much current research \citep{Marley93, Markham18, Wu19, Markham20}.
For example, \citet{Markham18} examined mode excitation mechanisms in giant planet interiors. They found that moist convective storms associated with water condensation were not energetically feasible for Jupiter, but it remains to be seen whether that mechanism could excite Uranian or Neptunian modes to observable levels.
Given all these theoretical uncertainties, we leave the computation of amplitudes for future work.

\subsection{Occultations of Narrow Dense Rings}

One way to identify resonant perturbations would be the detection of density waves similar to those seen in Saturn's rings. This would best work in Uranus's narrow dense rings, where the search can be guided by a combination of our location predictions and the range of the number of spiral arms $m$ corresponding to an approximation of the number of stellar occultation cuts through the rings or images of the rings that would be necessary for detection. 
Note that the resolution achievable with an occultation is limited by the Fresnel scale: $\lambda_F \sim \sqrt{\lambda D}$. At $\lambda = 0.2 \mu$m, the great distances $D$ of Uranus (19.2 AU) and Neptune (30 AU) from Earth prevent ground-based or Earth-orbit-based stellar occultations from obtaining better resolution than about 750 m and 950 m, respectively.
Given that wave-like variations in the dense rings have sub-kilometer wavelengths \citep{Chancia16, Horn88}, we expect resolutions of order 200 m is needed to see any planet-generated waves. This resolution threshold, combined with the number of cuts required to uniquely determine the number of arms, which is of order the relevant $m$ (about 20 for the waves of interest here), implies that the detection of planetary normal modes via wave identification will have to await an orbiter of Uranus or Neptune. 
An orbiter can observe occultations in the ultraviolet, visible, near-infrared, and radio parts of the spectrum at a much closer distance, and it would be expected to produce radial profiles of the rings at a range of longitudes with sufficient resolution to determine the number of spiral arms in a wave. Observation of the ring plane from different incidence angles will also allow bending waves to be easily discerned from density waves. 


While the overall shapes of the Uranian rings are dominated by factors that can be attributed to free modes in the rings themselves, this does not exclude the possibility that these rings can also preserve signals from planetary normal modes.
The Uranian $\alpha$ and $\beta$ rings are considered analogs of the Saturnian Maxwell and Colombo ringlets \citep{Porco90, Chiang03}.
These rings are considered in \citet{Chiang03}, who present a proof, relying on first steps taken by \citet{Borderies83}, that circular, nodally locked rings are linearly stable to perturbations to their inclinations and nodes.
The Colombo ringlet, also named the Titan ringlet because it is in a 1:0 apsidal resonance with Saturn's moon Titan \citep{Porco84, Nicholson14}, is located near several planetary normal modes: $6 \leq \ell \leq 15$, $5 \leq m \leq 11$ \citep{Mankovich19}.
The Maxwell ringlet inside the Maxwell Gap in Saturn's C ring is perturbed by an $\ell=m=2$ mode \citep{French16, Cuzzi18}, as predicted by \citet{Fuller14}. 
For this reason, we are also hopeful that either the $\alpha$ or $\beta$ rings, or both, are likewise perturbed by planetary normal modes. 
The Uranian $\epsilon$ ring is also considered to have similar properties to the Maxwell ringlet \citep{French91, French16}: sharp edges, a freely precessing elliptical shape, and a linear width-radius relation. 
The mean optical depth of the Maxwell ringlet is $\bar{\tau} = 0.968$ \citep{French16}. 

\citet{Horn88} identified an inward-propagating density wave in the $\delta$ ring, which they interpreted to be evidence of a moonlet interior to the $\delta$ ring. Because no moonlet has yet been discovered interior to the $\delta$ ring, we consider that such a density wave could be a candidate planetary normal mode resonance. 
\citet{Horn88} constrained the azimuthal wavenumber $m$ of the resonance that generated this wave to be $48 \leq m \leq 112$, based on three conditions: that the torque exceed a critical value \citep{Goldreich87} for nonlinearity and the moonlet still remain undetected in the Voyager Imaging search \citep{Smith86}, that the moonlet lie between the $\gamma$ and $\delta$ rings, and that the separation in radius between first-order resonances be greater than half the width of the $\delta$ ring \citep{Horn88}.
The range of possible $m$ values was thus not constrained by an $m$-lobed pattern detected in the ring itself.
If instead the inward-propagating density wave in the $\delta$ ring is driven by a planetary normal mode resonance, none of the conditions given by \citet{Horn88} would apply, though their measurement of the product of the wavelength and the distance from the resonance $\lambda d = 0.84 \pm 0.07$ km \citep{Horn88} makes it difficult for $m$ to be less than 10.
From our resonance location predictions, we can instead expect $17 \leq m \leq 23$, should the density wave in the $\delta$ ring be from an f-mode resonance. 

In addition to density waves, Lindblad resonances with satellites can perturb ring edges, generating non-circular shapes like those observed for the outer edge of Saturn's A and B rings \citep{Goldreich78, Nicholson14, ElMoutamid16, Tajeddine17}  and for Uranus's $\epsilon$ ring \citep{French91} and $\eta$ ring \citep{Chancia17}.

\subsection{Correlating Uranian Ring Features and Resonance Locations}

\begin{figure*}[htb] 
   \centering
   \includegraphics[width=\textwidth, trim={1.0in 0 1.0in 0}]{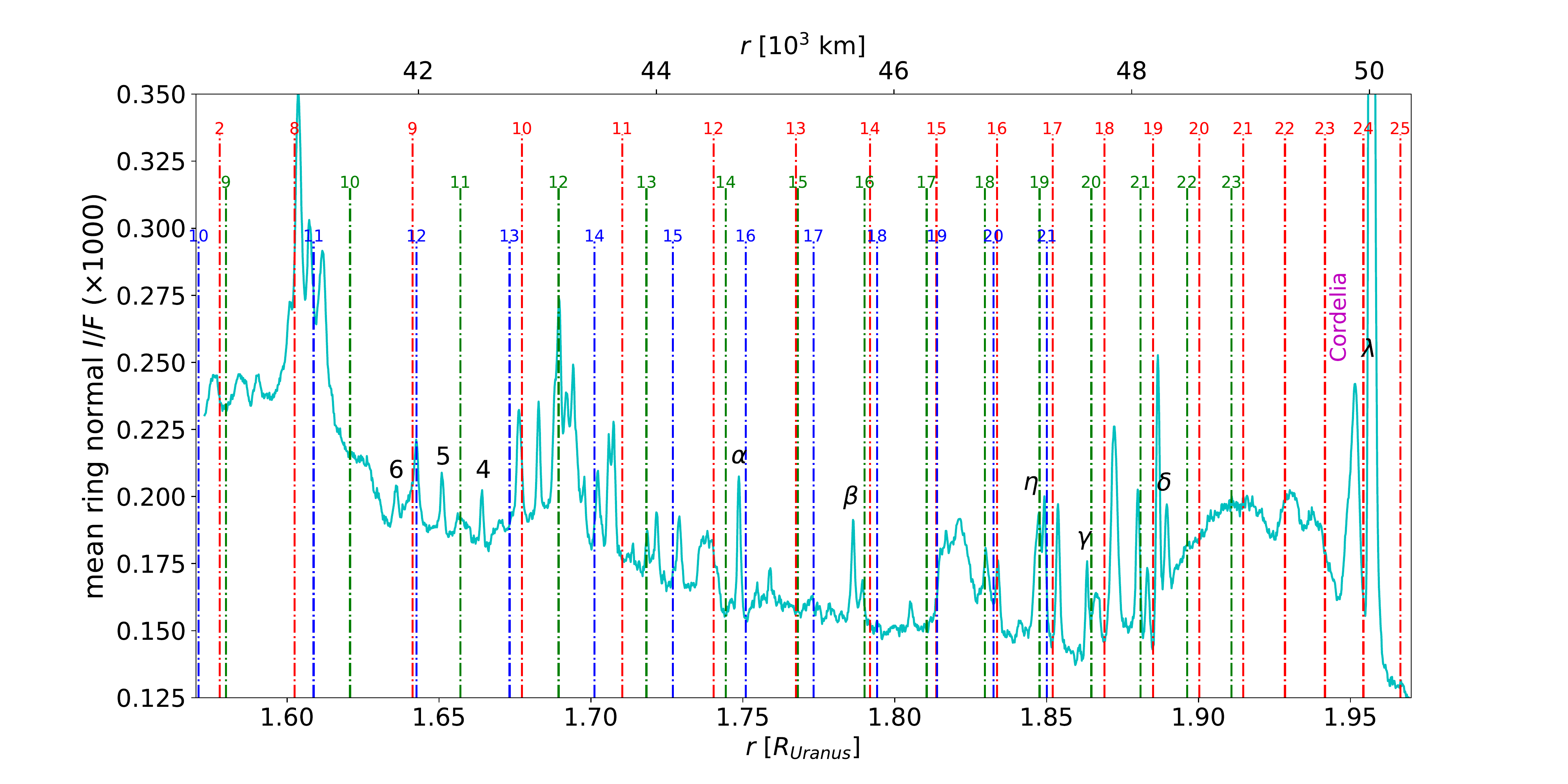} 
   \caption{Radial scan of Uranian rings compared to Lindblad resonance predictions from the medium model.
   The cyan curve shows the mean ring normal $I/F$ as a function of radius. 
   While the named narrow rings are labeled, many other narrow features can be seen in this scan.
   The tallest dash-dotted lines, in red, show the locations of the $\ell=m$ mode resonances; the intermediate-sized dash-dotted lines, in green, show the locations of the $\ell-m=2$ mode resonances, and the shortest dash-dotted lines, in blue, show the locations of the $\ell-m=4$ mode resonances. The integer above each dash-dotted line corresponds to the azimuthal order $m$ of the mode. 
   }
   \label{U_radscan}
\end{figure*}

Another potential way to identify these resonances would be to correlate the locations of multiple narrow ring features with expected resonant locations. 
For example, Figure \ref{U_radscan} shows the brightness of the inner rings of Uranus as a function of radius, from a high-phase image (C2685219; see \citealt{Hedman21}) that showed many more narrow ring features than the named rings. These features could potentially reflect additional locations where material is confined by resonances. 
Vertical dash-dotted lines show Lindblad resonance locations calculated from the medium model. 
Due to the uncertainty of our calculations, lining up ring features with resonances in this way is to be taken only as a demonstration of a way to correlate the models with ring features. 
Nevertheless, the way that the $\ell=m=8,9,10,11,12$ modes line up with either peaks or troughs in the ring brightness is suggestive of a role they may play in perturbing ring material radially. 

\subsection{Longitudinal Variations in Diffuse Rings}
The Uranian $\zeta$ ring and the Neptunian Galle ring have low optical depth. 
Although many mode resonances fall within them,
these rings are so tenuous that we do not expect them to be capable of sustaining a wave.
Longitudinal structure can be driven by resonances with the planet's magnetic field, as is the case at magnetic Lindblad resonances in Saturn's D Ring \citep{Hedman09b, Chancia19}, and we encourage future work exploring potential resonances between the Uranian and Neptunian magnetic fields and their ring systems. Nonetheless, planetary normal modes, particularly the stronger $\ell=m=2$ modes, should be kept in mind in studies of longitudinal structures in diffuse rings.

\subsection{Neptune's Le Verrier and Adams Rings \label{nep_lev_adams}}

Finally, we can consider whether planetary normal mode resonances could influence the dynamics of Neptune's narrow dusty rings in a detectable way.

While normal mode resonances could potentially play a role in confining the Le Verrier ring \citep{Brooks21}, testing this idea is challenging. 
For one, the ring appears homogeneous in the Voyager images \citep{Ferrari94}, which limits our ability to identify signatures of external perturbations. 
Furthermore, this ring is also close to other resonances, including Thalassa's 21:23 resonance \citep{Gaslac20}.
In addition, the 2:1 resonance with Neptune's rotation frequency, whose value is known with less certainty, falls in the Le Verrier ring region. Simply applying Kepler's Third Law to find the 2:1 resonance location, neglecting Neptune's oblateness, yields

\begin{equation}
    a_{\rm{res}} = \left( \frac{GM}{n^2} \right)^{1/3} = \begin{cases}
      52,607 \:\rm{km}, & \text{fast} \\
      55,507 \:\rm{km}, & \text{slow} 
    \end{cases} 
\end{equation}
where the mean motion $n = 2 \Omega$, and ``fast'' and ``slow'' refer to estimates for Neptune's rotation rate given in Table \ref{model_params}. The Le Verrier ring is located at 53,200 km \citep{DePater18}, in the middle of the resonance locations according to the two different rotation rates \citep{Warwick89, Helled10}.

It is also worth noting that $n=1$ g-mode resonances can occur near Neptune's Adams ring, because this could potentially mean that these resonances may be relevant to understanding the arcs in that ring.
The ring arcs in the Adams ring comprise the most studied part of Neptune's rings (see for example \citealt{Porco95, DePater18}). 
Ring arcs in the Saturnian system are confined longitudinally by orbital resonances with moons \citep{Spitale06, Hedman07, Hedman09, Hedman10, Cooper08, A19}.
Although several ideas have been proposed about the particular resonance of Neptune's ring arcs, consensus has not been reached. 
Initial studies suspected that a 42:43 resonance with Galatea confined the ring arcs \citep{Goldreich86, Porco91, Namouni02}. Deviations from the exact rate of different types of a corotation resonance with Galatea, however, support other possible explanations \citep{Renner14}.
Confinement due to shepherding by undetected satellites that are co-orbital with the Adams ring arcs has been proposed \citep{Lissauer85, Salo98, Renner14},
though more recent investigation claims to rule out that co-orbital satellites could be the source of the dust \citep{GW20}.
Another recent idea is that the Adams ring arcs are in a three-body resonance with Galatea and Larissa \citep{Showalter17}.

One potential way to test whether planetary normal modes could be confining material in the Adams ring is to take another look at the distribution and extent of the arcs.
A consequence of the initial 42:43 corotation resonance theory of confinement was that each of the 42 corotation sites, also called ``pockets'', spans only $\sim9^{\circ}$. This posed a problem for the Fraternit\'e arc, whose longitudinal extent is greater, and so it was assumed that Fraternit\'e occupies two corotation sites. 

Should the ring arcs instead be the effect of a corotation resonance with an oscillation degree $m < 42$ , the corotation site would span a greater longitudinal extent $\theta$, because

\begin{equation}
\theta = \frac{360^{\circ}}{m}, 
\end{equation}
which for low enough $m$ could encompass the entirety of Fraternit\'e in one corotation site. 
Although Lindblad resonances do not confine material, each Lindblad resonance can be associated with a corotation resonance, which can confine material.
The planetary corotation resonance from the $\ell=m=19$, $n=1$ g-mode is calculated to fall in the range $62,844$-$63,230$ km, which encompasses the Adams ring. 
The $m=19$ corotation site would span approximately  $19^{\circ}$, over twice as much as an $m=42$ corotation site, and could thus encompass the entirety of the Fraternit\'e arc. 
We hope to explore this possibility further in a future work. 

\subsection{Conclusion}

While none of the structures in the rings around Uranus and Neptune has yet been firmly attributed to a planetary normal mode, the above considerations indicate that ring seismology of the ice giants could be done when the appropriate data becomes available.
To the extent that different interior models affect normal mode resonance locations, the detection of planet-driven ring waves or perturbations to moon orbits would allow a variety of interior models to be ruled out.
These may provide evidence for or against a stably stratified layer or a diffuse core. 
Stable stratification, which occurs in all our interior models except the adiabatic model, allows the presence of g-modes, which could possibly be detected if they fall in or near rings \citep{Fuller14, Friedson20}. 
Should Uranus or Neptune additionally have a diffuse core, for example, the f-mode and g-mode resonances would be modified compared to our results here, particularly those corresponding to the $\ell=2$ modes most sensitive to $\rho(r)$ and $N(r)$ in the deep interior.

To evaluate these predictions, attempts can be made with Voyager and ground-based observations. Nevertheless, we expect results from such observations to be inconclusive, and an orbiter with a primary mission duration of at least a year or two to measure pattern speeds to within around $0.1^{\circ}$/day would likely be required to make the observations necessary to determine effects from planetary normal mode resonances in the rings of Uranus and Neptune. \\
\vspace{0.25in}
\\
We are grateful to many individuals for useful comments and discussions, including P. Nicholson, M. Tiscareno, M. \'Cuk, L. Fletcher, M. Brozovi\'c, and K. Volk. 
We thank two anonymous reviewers for their helpful comments that
have improved this paper.
We also thank NASA for the support through the Solar System Workings grants NNX15AH45G and 80NSSC21K1865. 

\bibliography{Bibliography} 
\bibliographystyle{aasjournal}

\appendix

\section{Extended Tables of f-mode Frequencies and Resonance Locations}
In this appendix, we present all the calculated f-mode frequencies and resonance locations for all the models. Table \ref{Marley_predictions} shows the initial estimates by \citet{Marley88}.
Our Lindblad resonance calculations are provided in tables that contains two models each: the thick and medium models in Table \ref{Lind_thick_med}, the thin and adiabatic models in Table \ref{Lind_thin_ad}, and the shallow and Neptune models in Table \ref{Lind_shallow_Nep}.
Similarly, our vertical resonance calculations are then provided in tables that contains two models each: the thick and medium models in Table \ref{vert_thick_med}, the thin and adiabatic models in Table \ref{vert_thin_ad}, and the shallow and Neptune models in Table \ref{vert_shallow_Nep}.
Error bars are one-sided because the frequencies and resonance locations are only calculated to first order. Second-order calculations would universally lower the frequencies, which would cause the resonance locations to be more distant from the planet. 

\begin{table}[!h]
\centering
\caption{Predicted Lindblad resonance locations among the inner rings of Uranus from \citet{Marley88}.
\label{Marley_predictions} }
\begin{tabular}{rrrrr p{9cm}}
\hline
\toprule
$\ell$ &$m$ &model &$\Omega_{\rm{pat}}$ (deg/day) &$r_{\rm{res}}$ (km)  \\

2  &2  &6 &1683 &48,300 \\
   &   &5 &1718 &47,700 \\
   &   &4 &1767 &46,800 \\
3  &3  &6 &2015 &39,600 \\
   &   &5 &2025 &39,500 \\
   &   &4 &2054 &39,100 \\
4  &4  &6 &2005 &38,100 \\
   &   &5 &2010 &38,000 \\
   &   &4 &2030 &37,800 \\
5  &5  &6 &1941 &37,900 \\
   &   &5 &1941 &37,900 \\
   &   &4 &1950 &37,800 \\
\hline
\end{tabular}
\end{table}

\begin{table}
\centering
\caption{Pattern frequencies and Lindblad resonance locations of different models
}
\renewcommand{\arraystretch}{0.5}
\begin{tabular}{@{}rrcc@{}}
\multicolumn{4}{c}{\textbf{Uranus Thick model}} \\
\toprule
$\ell$ &$m$ &$\Omega_{\rm{pat}}$ (deg/day) &$r_{\rm{res}}$ (km)  \\
\midrule
$\ell=m$ \\
2 &2 &2218.7 - 62.4 &40196 + 772 \\
3 &3 &2228.9 - 59.9 &37054 + 678 \\
4 &4 &2077.7 - 53.6 &37198 + 653 \\
5 &5 &1922.5 - 47.8 &38123 + 644 \\
6 &6 &1788.3 - 42.9 &39263 + 641 \\
7 &7 &1677.2 - 39.1 &40418 + 639 \\
8 &8 &1586.7 - 35.9 &41503 + 638 \\
9 &9 &1513.3 - 33.4 &42481 + 636 \\
10 &10 &1453.5 - 31.4 &43345 + 634 \\
11 &11 &1404.4 - 29.7 &44106 + 632 \\
12 &12 &1363.3 - 28.3 &44780 + 629 \\
13 &13 &1328.1 - 27.0 &45386 + 626 \\
14 &14 &1297.5 - 26.0 &45939 + 623 \\
15 &15 &1270.5 - 25.0 &46451 + 620 \\
16 &16 &1246.2 - 24.2 &46928 + 617 \\
17 &17 &1224.3 - 23.4 &47377 + 614 \\
18 &18 &1204.3 - 22.7 &47802 + 611 \\
19 &19 &1185.9 - 22.1 &48205 + 608 \\
20 &20 &1169.0 - 21.5 &48589 + 605 \\
21 &21 &1153.2 - 21.0 &48955 + 602 \\
22 &22 &1138.6 - 20.5 &49305 + 600 \\
23 &23 &1125.0 - 20.0 &49640 + 597 \\
24 &24 &1112.2 - 19.6 &49962 + 594 \\
25 &25 &1100.2 - 19.2 &50271 + 592 \\
\midrule
$\ell-m=2$ \\
4 &2 &3755.8 - 107.2 &28306 + 551 \\
5 &3 &2919.8 - 79.6 &30955 + 575 \\
6 &4 &2460.7 - 64.4 &33234 + 592 \\
7 &5 &2166.4 - 54.7 &35208 + 604 \\
8 &6 &1961.7 - 47.9 &36916 + 613 \\
9 &7 &1812.2 - 43.0 &38388 + 618 \\
10 &8 &1699.1 - 39.2 &39654 + 621 \\
11 &9 &1610.9 - 36.3 &40749 + 623 \\
12 &10 &1540.2 - 33.9 &41705 + 623 \\
13 &11 &1482.1 - 32.0 &42553 + 622 \\
14 &12 &1433.1 - 30.3 &43314 + 621 \\
15 &13 &1391.1 - 28.9 &44007 + 619 \\
16 &14 &1354.5 - 27.6 &44644 + 617 \\
17 &15 &1322.1 - 26.6 &45234 + 615 \\
18 &16 &1293.3 - 25.6 &45784 + 613 \\
19 &17 &1267.3 - 24.7 &46299 + 611 \\
20 &18 &1243.8 - 23.9 &46784 + 609 \\
21 &19 &1222.4 - 23.2 &47242 + 606 \\
22 &20 &1202.7 - 22.5 &47675 + 604 \\
23 &21 &1184.6 - 21.9 &48087 + 602 \\
24 &22 &1167.9 - 21.3 &48478 + 599 \\
25 &23 &1152.3 - 20.8 &48852 + 597 \\
\midrule
$\ell-m=4$ \\
7 &3 &3307.8 - 91.1 &28487 + 535 \\
8 &4 &2711.8 - 71.9 &31153 + 562 \\
9 &5 &2350.3 - 60.1 &33349 + 581 \\
10 &6 &2108.3 - 52.3 &35187 + 593 \\
11 &7 &1935.4 - 46.6 &36743 + 602 \\
12 &8 &1805.6 - 42.4 &38080 + 607 \\
13 &9 &1704.4 - 39.1 &39246 + 610 \\
14 &10 &1622.9 - 36.4 &40278 + 612 \\
15 &11 &1555.6 - 34.1 &41203 + 613 \\
16 &12 &1498.8 - 32.3 &42041 + 614 \\
17 &13 &1450.1 - 30.6 &42807 + 613 \\
18 &14 &1407.7 - 29.2 &43512 + 612 \\
19 &15 &1370.5 - 28.0 &44165 + 611 \\
20 &16 &1337.4 - 26.9 &44772 + 610 \\
21 &17 &1307.8 - 25.9 &45340 + 608 \\
22 &18 &1281.1 - 25.0 &45873 + 607 \\
23 &19 &1256.9 - 24.2 &46375 + 605 \\
24 &20 &1234.7 - 23.5 &46849 + 603 \\
25 &21 &1214.4 - 22.8 &47298 + 601 \\
\bottomrule                          
\end{tabular}
\begin{tabular}{@{}rrcc@{}}
\multicolumn{4}{c}{\textbf{Uranus Medium model}} \\
\toprule
$\ell$ &$m$ &$\Omega_{\rm{pat}}$ (deg/day) &$r_{\rm{res}}$ (km)  \\
\midrule
$\ell=m$ \\
2 &2 &2207.8 - 62.1 &40328 + 774 \\
3 &3 &2234.6 - 60.1 &36991 + 678 \\
4 &4 &2096.0 - 54.2 &36981 + 651 \\
5 &5 &1949.2 - 48.6 &37774 + 641 \\
6 &6 &1819.6 - 43.9 &38812 + 636 \\
7 &7 &1710.1 - 40.0 &39899 + 634 \\
8 &8 &1618.6 - 36.8 &40956 + 632 \\
9 &9 &1542.1 - 34.2 &41951 + 630 \\
10 &10 &1477.8 - 32.0 &42870 + 629 \\
11 &11 &1423.3 - 30.1 &43714 + 627 \\
12 &12 &1377.0 - 28.6 &44481 + 626 \\
13 &13 &1337.5 - 27.3 &45174 + 624 \\
14 &14 &1303.6 - 26.1 &45797 + 621 \\
15 &15 &1274.3 - 25.1 &46358 + 619 \\
16 &16 &1248.7 - 24.3 &46867 + 616 \\
17 &17 &1226.0 - 23.5 &47335 + 614 \\
18 &18 &1205.5 - 22.8 &47771 + 611 \\
19 &19 &1186.8 - 22.1 &48180 + 608 \\
20 &20 &1169.7 - 21.5 &48568 + 605 \\
21 &21 &1153.9 - 21.0 &48937 + 602 \\
22 &22 &1139.2 - 20.5 &49289 + 600 \\
23 &23 &1125.4 - 20.0 &49626 + 597 \\
24 &24 &1112.6 - 19.6 &49949 + 594 \\
25 &25 &1100.5 - 19.2 &50259 + 592 \\
\midrule
$\ell-m=2$ \\
4 &2 &3792.6 - 108.4 &28123 + 548 \\
5 &3 &2963.8 - 81.0 &30649 + 571 \\
6 &4 &2506.6 - 65.8 &32828 + 587 \\
7 &5 &2210.9 - 56.0 &34735 + 598 \\
8 &6 &2002.6 - 49.1 &36413 + 606 \\
9 &7 &1847.7 - 43.9 &37895 + 612 \\
10 &8 &1728.0 - 40.0 &39211 + 616 \\
11 &9 &1633.0 - 36.8 &40381 + 618 \\
12 &10 &1556.0 - 34.3 &41424 + 619 \\
13 &11 &1492.6 - 32.2 &42352 + 620 \\
14 &12 &1439.9 - 30.5 &43178 + 619 \\
15 &13 &1395.4 - 29.0 &43917 + 618 \\
16 &14 &1357.2 - 27.7 &44584 + 617 \\
17 &15 &1324.0 - 26.6 &45192 + 615 \\
18 &16 &1294.6 - 25.6 &45753 + 613 \\
19 &17 &1268.4 - 24.7 &46275 + 611 \\
20 &18 &1244.6 - 23.9 &46764 + 609 \\
21 &19 &1223.1 - 23.2 &47224 + 606 \\
22 &20 &1203.3 - 22.5 &47660 + 604 \\
23 &21 &1185.1 - 21.9 &48073 + 602 \\
24 &22 &1168.3 - 21.4 &48466 + 599 \\
25 &23 &1152.7 - 20.8 &48840 + 597 \\

\midrule
$\ell-m=4$ \\
7 &3 &3379.5 - 93.3 &28084 + 528 \\
8 &4 &2770.6 - 73.6 &30711 + 555 \\
9 &5 &2397.7 - 61.5 &32909 + 574 \\
10 &6 &2145.0 - 53.3 &34785 + 587 \\
11 &7 &1962.4 - 47.3 &36406 + 597 \\
12 &8 &1824.4 - 42.9 &37819 + 603 \\
13 &9 &1716.8 - 39.4 &39058 + 608 \\
14 &10 &1630.7 - 36.6 &40150 + 611 \\
15 &11 &1560.4 - 34.3 &41117 + 612 \\
16 &12 &1501.9 - 32.3 &41983 + 613 \\
17 &13 &1452.2 - 30.7 &42766 + 613 \\
18 &14 &1409.2 - 29.3 &43481 + 612 \\
19 &15 &1371.6 - 28.0 &44140 + 611 \\
20 &16 &1338.3 - 26.9 &44752 + 610 \\
21 &17 &1308.6 - 25.9 &45323 + 608 \\
22 &18 &1281.8 - 25.0 &45858 + 607 \\
23 &19 &1257.4 - 24.2 &46361 + 605 \\
24 &20 &1235.2 - 23.5 &46836 + 603 \\
25 &21 &1214.9 - 22.8 &47286 + 601 \\
\bottomrule                          
\end{tabular}
\label{Lind_thick_med}

\end{table}

\newpage 
\begin{table}
\centering
\caption{Pattern frequencies and Lindblad resonance locations of different models
}
\renewcommand{\arraystretch}{0.5}
\begin{tabular}{@{}rrcc@{}}
\multicolumn{4}{c}{\textbf{Uranus Thin model}} \\
\toprule
$\ell$ &$m$ &$\Omega_{\rm{pat}}$ (deg/day) &$r_{\rm{res}}$ (km)  \\
\midrule
$\ell=m$ \\
2 &2 &2191.7 - 61.6 &40525 + 777 \\
3 &3 &2235.0 - 60.1 &36986 + 678 \\
4 &4 &2108.1 - 54.6 &36840 + 650 \\
5 &5 &1968.8 - 49.3 &37524 + 639 \\
6 &6 &1843.2 - 44.7 &38481 + 634 \\
7 &7 &1735.2 - 40.8 &39514 + 631 \\
8 &8 &1643.7 - 37.6 &40538 + 629 \\
9 &9 &1566.4 - 34.9 &41517 + 627 \\
10 &10 &1500.6 - 32.6 &42435 + 626 \\
11 &11 &1444.4 - 30.7 &43288 + 624 \\
12 &12 &1395.9 - 29.1 &44080 + 622 \\
13 &13 &1353.6 - 27.6 &44814 + 620 \\
14 &14 &1316.7 - 26.4 &45494 + 618 \\
15 &15 &1284.2 - 25.3 &46121 + 616 \\
16 &16 &1255.6 - 24.4 &46695 + 615 \\
17 &17 &1230.6 - 23.6 &47216 + 612 \\
18 &18 &1208.5 - 22.8 &47691 + 610 \\
19 &19 &1188.9 - 22.2 &48125 + 608 \\
20 &20 &1171.2 - 21.6 &48528 + 605 \\
21 &21 &1155.0 - 21.0 &48906 + 602 \\
22 &22 &1140.0 - 20.5 &49264 + 600 \\
23 &23 &1126.2 - 20.0 &49605 + 597 \\
24 &24 &1113.2 - 19.6 &49931 + 594 \\
25 &25 &1101.1 - 19.2 &50243 + 592 \\
\midrule
$\ell-m=2$ \\
4 &2 &3818.5 - 109.2 &27996 + 547 \\
5 &3 &2997.4 - 82.1 &30419 + 568 \\
6 &4 &2542.4 - 67.0 &32520 + 583 \\
7 &5 &2246.1 - 57.1 &34371 + 594 \\
8 &6 &2035.9 - 50.1 &36015 + 602 \\
9 &7 &1878.4 - 44.8 &37482 + 608 \\
10 &8 &1756.0 - 40.8 &38794 + 612 \\
11 &9 &1658.0 - 37.5 &39974 + 614 \\
12 &10 &1577.8 - 34.9 &41040 + 615 \\
13 &11 &1511.0 - 32.7 &42008 + 616 \\
14 &12 &1454.5 - 30.8 &42888 + 616 \\
15 &13 &1406.3 - 29.2 &43690 + 616 \\
16 &14 &1364.8 - 27.9 &44418 + 615 \\
17 &15 &1329.0 - 26.7 &45077 + 614 \\
18 &16 &1297.9 - 25.7 &45675 + 613 \\
19 &17 &1270.6 - 24.8 &46220 + 611 \\
20 &18 &1246.2 - 24.0 &46724 + 609 \\
21 &19 &1224.3 - 23.2 &47194 + 606 \\
22 &20 &1204.3 - 22.6 &47635 + 604 \\
23 &21 &1185.9 - 22.0 &48052 + 602 \\
24 &22 &1169.0 - 21.4 &48447 + 599 \\
25 &23 &1153.3 - 20.9 &48824 + 597 \\
\midrule
$\ell-m=4$ \\
7 &3 &3438.3 - 95.2 &27763 + 524 \\
8 &4 &2820.2 - 75.1 &30350 + 551 \\
9 &5 &2440.1 - 62.8 &32527 + 569 \\
10 &6 &2181.5 - 54.4 &34396 + 583 \\
11 &7 &1993.7 - 48.2 &36024 + 592 \\
12 &8 &1850.8 - 43.6 &37459 + 599 \\
13 &9 &1738.3 - 39.9 &38734 + 604 \\
14 &10 &1647.5 - 37.0 &39876 + 607 \\
15 &11 &1572.8 - 34.6 &40902 + 610 \\
16 &12 &1510.4 - 32.5 &41825 + 611 \\
17 &13 &1457.8 - 30.8 &42656 + 612 \\
18 &14 &1412.9 - 29.4 &43405 + 612 \\
19 &15 &1374.1 - 28.1 &44087 + 611 \\
20 &16 &1340.1 - 27.0 &44713 + 610 \\
21 &17 &1309.9 - 26.0 &45292 + 608 \\
22 &18 &1282.8 - 25.1 &45833 + 607 \\
23 &19 &1258.3 - 24.3 &46340 + 605 \\
24 &20 &1236.0 - 23.5 &46818 + 603 \\
25 &21 &1215.5 - 22.8 &47270 + 601 \\
\bottomrule                          
\end{tabular}
\begin{tabular}{@{}rrcc@{}}
\multicolumn{4}{c}{\textbf{Uranus Adiabatic model}} \\
\toprule
$\ell$ &$m$ &$\Omega_{\rm{pat}}$ (deg/day) &$r_{\rm{res}}$ (km)  \\
\midrule
$\ell=m$ \\
2 &2 &2360.4 - 67.0 &38571 + 747 \\
3 &3 &2150.9 - 57.5 &37943 + 692 \\
4 &4 &1963.4 - 50.2 &38627 + 672 \\
5 &5 &1820.2 - 44.8 &39537 + 662 \\
6 &6 &1710.1 - 40.7 &40451 + 655 \\
7 &7 &1623.0 - 37.6 &41314 + 649 \\
8 &8 &1552.2 - 35.0 &42115 + 645 \\
9 &9 &1493.3 - 32.9 &42858 + 640 \\
10 &10 &1443.4 - 31.1 &43547 + 637 \\
11 &11 &1400.4 - 29.6 &44189 + 633 \\
12 &12 &1362.9 - 28.3 &44787 + 630 \\
13 &13 &1329.8 - 27.1 &45349 + 626 \\
14 &14 &1300.2 - 26.1 &45876 + 623 \\
15 &15 &1273.7 - 25.1 &46373 + 620 \\
16 &16 &1249.6 - 24.3 &46843 + 617 \\
17 &17 &1227.7 - 23.5 &47289 + 614 \\
18 &18 &1207.7 - 22.8 &47712 + 611 \\
19 &19 &1189.3 - 22.2 &48115 + 608 \\
20 &20 &1172.2 - 21.6 &48499 + 605 \\
21 &21 &1156.4 - 21.1 &48866 + 602 \\
22 &22 &1141.7 - 20.6 &49217 + 600 \\
23 &23 &1127.9 - 20.1 &49553 + 597 \\
24 &24 &1115.0 - 19.7 &49876 + 594 \\
25 &25 &1102.9 - 19.2 &50187 + 592 \\
\midrule
$\ell-m=2$ \\
4 &2 &3534.8 - 100.4 &29473 + 571 \\
5 &3 &2755.2 - 74.6 &32175 + 594 \\
6 &4 &2347.7 - 61.1 &34291 + 607 \\
7 &5 &2093.3 - 52.6 &36022 + 615 \\
8 &6 &1917.5 - 46.7 &37481 + 620 \\
9 &7 &1787.6 - 42.3 &38739 + 623 \\
10 &8 &1687.0 - 38.9 &39843 + 624 \\
11 &9 &1606.3 - 36.2 &40826 + 624 \\
12 &10 &1539.9 - 33.9 &41711 + 623 \\
13 &11 &1484.1 - 32.0 &42515 + 622 \\
14 &12 &1436.3 - 30.4 &43251 + 621 \\
15 &13 &1394.8 - 29.0 &43930 + 619 \\
16 &14 &1358.3 - 27.8 &44559 + 617 \\
17 &15 &1326.0 - 26.7 &45145 + 615 \\
18 &16 &1297.1 - 25.7 &45694 + 613 \\
19 &17 &1271.1 - 24.8 &46209 + 611 \\
20 &18 &1247.4 - 24.0 &46694 + 609 \\
21 &19 &1225.9 - 23.3 &47153 + 606 \\
22 &20 &1206.1 - 22.6 &47587 + 604 \\
23 &21 &1187.9 - 22.0 &48000 + 602 \\
24 &22 &1171.0 - 21.4 &48392 + 599 \\
25 &23 &1155.3 - 20.9 &48767 + 597 \\
\midrule
$\ell-m=4$ \\
6 &2 &4260.5 - 122.2 &26026 + 509 \\
7 &3 &3190.8 - 87.6 &29179 + 546 \\
8 &4 &2648.1 - 70.0 &31649 + 570 \\
9 &5 &2317.2 - 59.2 &33665 + 585 \\
10 &6 &2092.9 - 51.9 &35359 + 596 \\
11 &7 &1929.9 - 46.5 &36813 + 603 \\
12 &8 &1805.5 - 42.4 &38083 + 607 \\
13 &9 &1706.9 - 39.1 &39208 + 610 \\
14 &10 &1626.7 - 36.5 &40215 + 612 \\
15 &11 &1559.9 - 34.3 &41126 + 613 \\
16 &12 &1503.3 - 32.4 &41957 + 613 \\
17 &13 &1454.6 - 30.8 &42719 + 613 \\
18 &14 &1412.1 - 29.4 &43422 + 612 \\
19 &15 &1374.7 - 28.1 &44074 + 611 \\
20 &16 &1341.5 - 27.0 &44682 + 610 \\
21 &17 &1311.7 - 26.0 &45251 + 608 \\
22 &18 &1284.8 - 25.1 &45785 + 607 \\
23 &19 &1260.4 - 24.3 &46288 + 605 \\
24 &20 &1238.2 - 23.6 &46763 + 603 \\
25 &21 &1217.7 - 22.9 &47212 + 601 \\

\bottomrule                          
\end{tabular}
\label{Lind_thin_ad}

\end{table}

\newpage 
\begin{table}
\centering
\caption{Pattern frequencies and Lindblad resonance locations of different models
}
\renewcommand{\arraystretch}{0.5}
\begin{tabular}{@{}rrcc@{}}
\multicolumn{4}{c}{\textbf{Uranus Shallow model}} \\
\toprule
$\ell$ &$m$ &$\Omega_{\rm{pat}}$ (deg/day) &$r_{\rm{res}}$ (km)  \\
\midrule
$\ell=m$ \\
2 &2 &2052.5 - 57.2 &42337 + 805 \\
3 &3 &2162.1 - 57.9 &37812 + 690 \\
4 &4 &2073.0 - 53.7 &37254 + 657 \\
5 &5 &1959.5 - 49.2 &37642 + 643 \\
6 &6 &1851.0 - 45.1 &38373 + 636 \\
7 &7 &1753.2 - 41.6 &39244 + 633 \\
8 &8 &1666.7 - 38.5 &40166 + 631 \\
9 &9 &1591.0 - 35.9 &41088 + 629 \\
10 &10 &1525.1 - 33.6 &41981 + 628 \\
11 &11 &1467.8 - 31.6 &42827 + 626 \\
12 &12 &1418.1 - 30.0 &43618 + 625 \\
13 &13 &1375.0 - 28.5 &44350 + 623 \\
14 &14 &1337.3 - 27.2 &45024 + 621 \\
15 &15 &1304.3 - 26.1 &45644 + 618 \\
16 &16 &1275.3 - 25.1 &46214 + 616 \\
17 &17 &1249.4 - 24.2 &46741 + 613 \\
18 &18 &1226.3 - 23.4 &47229 + 611 \\
19 &19 &1205.4 - 22.7 &47685 + 608 \\
20 &20 &1186.4 - 22.1 &48112 + 605 \\
21 &21 &1169.0 - 21.5 &48514 + 603 \\
22 &22 &1153.0 - 20.9 &48895 + 600 \\
23 &23 &1138.2 - 20.4 &49256 + 598 \\
24 &24 &1124.3 - 20.0 &49601 + 595 \\
25 &25 &1111.5 - 19.5 &49930 + 593 \\
\midrule
$\ell-m=2$ \\
4 &2 &3753.5 - 107.3 &28318 + 553 \\
5 &3 &2986.2 - 82.0 &30495 + 570 \\
6 &4 &2557.7 - 67.7 &32390 + 584 \\
7 &5 &2274.2 - 58.2 &34088 + 594 \\
8 &6 &2068.9 - 51.4 &35631 + 602 \\
9 &7 &1912.1 - 46.1 &37041 + 607 \\
10 &8 &1788.2 - 42.0 &38327 + 612 \\
11 &9 &1688.0 - 38.7 &39500 + 614 \\
12 &10 &1605.7 - 35.9 &40565 + 616 \\
13 &11 &1537.1 - 33.7 &41532 + 617 \\
14 &12 &1479.3 - 31.7 &42409 + 617 \\
15 &13 &1430.0 - 30.1 &43206 + 616 \\
16 &14 &1387.5 - 28.7 &43933 + 615 \\
17 &15 &1350.5 - 27.4 &44598 + 614 \\
18 &16 &1318.0 - 26.4 &45211 + 612 \\
19 &17 &1289.1 - 25.4 &45778 + 611 \\
20 &18 &1263.2 - 24.5 &46306 + 609 \\
21 &19 &1239.8 - 23.7 &46799 + 606 \\
22 &20 &1218.5 - 23.0 &47263 + 604 \\
23 &21 &1199.1 - 22.4 &47701 + 602 \\
24 &22 &1181.1 - 21.8 &48115 + 600 \\
25 &23 &1164.6 - 21.2 &48508 + 598 \\

\midrule
$\ell-m=4$ \\
7 &3 &3490.0 - 97.1 &27489 + 521 \\
8 &4 &2873.3 - 77.1 &29975 + 547 \\
9 &5 &2490.0 - 64.6 &32091 + 566 \\
10 &6 &2226.7 - 56.0 &33930 + 580 \\
11 &7 &2034.1 - 49.7 &35547 + 591 \\
12 &8 &1887.1 - 44.9 &36978 + 598 \\
13 &9 &1771.4 - 41.1 &38252 + 603 \\
14 &10 &1678.1 - 38.1 &39391 + 607 \\
15 &11 &1601.4 - 35.6 &40414 + 609 \\
16 &12 &1537.2 - 33.5 &41338 + 610 \\
17 &13 &1482.8 - 31.7 &42176 + 611 \\
18 &14 &1435.9 - 30.1 &42942 + 611 \\
19 &15 &1395.1 - 28.8 &43645 + 610 \\
20 &16 &1359.2 - 27.6 &44294 + 609 \\
21 &17 &1327.2 - 26.5 &44897 + 608 \\
22 &18 &1298.6 - 25.6 &45460 + 606 \\
23 &19 &1272.8 - 24.7 &45988 + 605 \\
24 &20 &1249.3 - 23.9 &46484 + 603 \\
25 &21 &1227.9 - 23.2 &46953 + 601 \\
\bottomrule                          
\end{tabular}
\begin{tabular}{@{}rrcc@{}}
\multicolumn{4}{c}{\textbf{Neptune model}} \\
\toprule
$\ell$ &$m$ &$\Omega_{\rm{pat}}$ (deg/day) &$r_{\rm{res}}$ (km)  \\
\midrule
$\ell=m$ \\
2 &2 &2292.5 - 45.4 &41555 + 557 \\
3 &3 &2251.9 - 42.6 &38883 + 498 \\
4 &4 &2088.3 - 38.1 &39168 + 483 \\
5 &5 &1940.9 - 34.3 &40024 + 478 \\
6 &6 &1821.9 - 31.3 &40973 + 475 \\
7 &7 &1726.4 - 28.9 &41890 + 474 \\
8 &8 &1648.6 - 27.0 &42747 + 472 \\
9 &9 &1583.8 - 25.4 &43541 + 471 \\
10 &10 &1529.0 - 24.1 &44279 + 470 \\
11 &11 &1481.7 - 22.9 &44966 + 469 \\
12 &12 &1440.4 - 21.9 &45608 + 468 \\
13 &13 &1404.0 - 21.0 &46211 + 467 \\
14 &14 &1371.5 - 20.2 &46779 + 466 \\
15 &15 &1342.2 - 19.5 &47315 + 465 \\
16 &16 &1315.7 - 18.9 &47823 + 463 \\
17 &17 &1291.5 - 18.3 &48306 + 462 \\
18 &18 &1269.4 - 17.8 &48765 + 461 \\
19 &19 &1249.0 - 17.3 &49203 + 460 \\
20 &20 &1230.1 - 16.9 &49622 + 459 \\
21 &21 &1212.6 - 16.5 &50023 + 457 \\
22 &22 &1196.3 - 16.1 &50407 + 456 \\
23 &23 &1181.0 - 15.7 &50776 + 455 \\
24 &24 &1166.7 - 15.4 &51131 + 454 \\
25 &25 &1153.3 - 15.1 &51472 + 453 \\
\midrule
$\ell-m=2$ \\
4 &2 &3803.1 - 76.1 &29659 + 402 \\
5 &3 &2969.9 - 57.1 &32337 + 421 \\
6 &4 &2526.1 - 46.9 &34505 + 433 \\
7 &5 &2247.0 - 40.5 &36304 + 442 \\
8 &6 &2053.6 - 36.0 &37832 + 448 \\
9 &7 &1910.6 - 32.6 &39154 + 452 \\
10 &8 &1799.8 - 30.1 &40319 + 455 \\
11 &9 &1711.0 - 28.0 &41358 + 457 \\
12 &10 &1637.8 - 26.3 &42297 + 458 \\
13 &11 &1576.2 - 24.8 &43152 + 459 \\
14 &12 &1523.4 - 23.6 &43938 + 460 \\
15 &13 &1477.6 - 22.5 &44664 + 460 \\
16 &14 &1437.4 - 21.6 &45338 + 460 \\
17 &15 &1401.6 - 20.8 &45969 + 460 \\
18 &16 &1369.6 - 20.0 &46560 + 459 \\
19 &17 &1340.8 - 19.4 &47116 + 459 \\
20 &18 &1314.6 - 18.7 &47642 + 458 \\
21 &19 &1290.6 - 18.2 &48140 + 457 \\
22 &20 &1268.7 - 17.7 &48612 + 457 \\
23 &21 &1248.4 - 17.2 &49062 + 456 \\
24 &22 &1229.7 - 16.8 &49491 + 455 \\
25 &23 &1212.2 - 16.4 &49901 + 454 \\
\midrule
$\ell-m=4$ \\
6 &2 &4638.9 - 93.8 &25983 + 356 \\
7 &3 &3461.9 - 67.4 &29198 + 385 \\
8 &4 &2863.8 - 54.0 &31739 + 405 \\
9 &5 &2498.8 - 45.7 &33825 + 418 \\
10 &6 &2251.3 - 40.1 &35586 + 428 \\
11 &7 &2071.3 - 36.0 &37104 + 436 \\
12 &8 &1933.8 - 32.9 &38436 + 441 \\
13 &9 &1824.9 - 30.4 &39620 + 445 \\
14 &10 &1736.2 - 28.3 &40684 + 448 \\
15 &11 &1662.3 - 26.6 &41650 + 451 \\
16 &12 &1599.6 - 25.2 &42533 + 452 \\
17 &13 &1545.6 - 24.0 &43345 + 454 \\
18 &14 &1498.5 - 22.9 &44097 + 454 \\
19 &15 &1457.0 - 21.9 &44797 + 455 \\
20 &16 &1420.1 - 21.1 &45451 + 455 \\
21 &17 &1387.0 - 20.3 &46064 + 455 \\
22 &18 &1357.2 - 19.6 &46641 + 455 \\
23 &19 &1330.0 - 19.0 &47186 + 455 \\
24 &20 &1305.2 - 18.4 &47702 + 454 \\
25 &21 &1282.4 - 17.9 &48191 + 454 \\
\bottomrule                          
\end{tabular}
\label{Lind_shallow_Nep}

\end{table}

\newpage 
\begin{table}
\centering
\caption{Pattern frequencies and vertical resonance locations of different models
}
\renewcommand{\arraystretch}{0.5}
\begin{tabular}{@{}rrcc@{}}
\multicolumn{4}{c}{\textbf{Uranus Thick model}} \\
\toprule
$\ell$ &$m$ &$\Omega_{\rm{pat}}$ (deg/day) &$r_{\rm{res}}$ (km)  \\
\midrule
$\ell-m=1$ \\
3 &2 &3166.5 - 89.8 &31763 + 613 \\
4 &3 &2637.1 - 71.5 &33162 + 611 \\
5 &4 &2296.5 - 59.7 &34825 + 615 \\
6 &5 &2057.3 - 51.5 &36463 + 621 \\
7 &6 &1881.0 - 45.6 &37981 + 625 \\
8 &7 &1747.4 - 41.1 &39344 + 628 \\
9 &8 &1644.1 - 37.6 &40545 + 629 \\
10 &9 &1562.7 - 34.8 &41593 + 629 \\
11 &10 &1497.3 - 32.6 &42507 + 628 \\
12 &11 &1443.7 - 30.8 &43311 + 627 \\
13 &12 &1398.7 - 29.3 &44029 + 625 \\
14 &13 &1360.1 - 28.0 &44680 + 623 \\
15 &14 &1326.5 - 26.8 &45276 + 620 \\
16 &15 &1296.7 - 25.8 &45828 + 618 \\
17 &16 &1270.2 - 24.9 &46344 + 615 \\
18 &17 &1246.2 - 24.1 &46827 + 612 \\
19 &18 &1224.4 - 23.3 &47284 + 610 \\
20 &19 &1204.4 - 22.7 &47715 + 607 \\
21 &20 &1186.1 - 22.0 &48125 + 605 \\
22 &21 &1169.2 - 21.5 &48514 + 602 \\
23 &22 &1153.4 - 20.9 &48886 + 599 \\
24 &23 &1138.8 - 20.4 &49241 + 597 \\
25 &24 &1125.2 - 20.0 &49581 + 594 \\
\midrule
$\ell-m=3$ \\
5 &2 &4166.5 - 119.4 &26475 + 515 \\
6 &3 &3133.2 - 85.9 &29574 + 551 \\
7 &4 &2594.4 - 68.3 &32113 + 575 \\
8 &5 &2261.8 - 57.5 &34235 + 591 \\
9 &6 &2036.4 - 50.1 &36027 + 602 \\
10 &7 &1874.5 - 44.8 &37549 + 609 \\
11 &8 &1752.9 - 40.8 &38853 + 614 \\
12 &9 &1658.2 - 37.7 &39983 + 616 \\
13 &10 &1582.1 - 35.1 &40977 + 617 \\
14 &11 &1519.4 - 33.1 &41863 + 618 \\
15 &12 &1466.5 - 31.3 &42664 + 617 \\
16 &13 &1421.1 - 29.8 &43394 + 616 \\
17 &14 &1381.5 - 28.5 &44067 + 615 \\
18 &15 &1346.7 - 27.3 &44689 + 613 \\
19 &16 &1315.7 - 26.2 &45269 + 612 \\
20 &17 &1287.9 - 25.3 &45812 + 610 \\
21 &18 &1262.7 - 24.5 &46322 + 608 \\
22 &19 &1239.9 - 23.7 &46803 + 606 \\
23 &20 &1218.9 - 23.0 &47257 + 603 \\
24 &21 &1199.7 - 22.4 &47688 + 601 \\
25 &22 &1182.0 - 21.8 &48097 + 599 \\
\midrule
$\ell-m=5$ \\
8 &3 &3461.8 - 95.8 &27679 + 520 \\
9 &4 &2821.1 - 75.2 &30374 + 550 \\
10 &5 &2435.7 - 62.7 &32590 + 570 \\
11 &6 &2178.8 - 54.4 &34444 + 584 \\
12 &7 &1995.2 - 48.4 &36022 + 594 \\
13 &8 &1857.3 - 43.9 &37386 + 600 \\
14 &9 &1749.4 - 40.4 &38583 + 605 \\
15 &10 &1662.5 - 37.6 &39648 + 608 \\
16 &11 &1590.6 - 35.2 &40606 + 609 \\
17 &12 &1530.0 - 33.2 &41476 + 610 \\
18 &13 &1478.1 - 31.5 &42271 + 610 \\
19 &14 &1433.1 - 30.0 &43004 + 610 \\
20 &15 &1393.6 - 28.7 &43683 + 609 \\
21 &16 &1358.5 - 27.5 &44314 + 608 \\
22 &17 &1327.2 - 26.5 &44904 + 607 \\
23 &18 &1299.0 - 25.6 &45457 + 606 \\
24 &19 &1273.4 - 24.7 &45977 + 604 \\
25 &20 &1250.1 - 23.9 &46469 + 602 \\
\bottomrule                          
\end{tabular}
\begin{tabular}{@{}rrcc@{}}
\multicolumn{4}{c}{\textbf{Uranus Medium model}} \\
\toprule
$\ell$ &$m$ &$\Omega_{\rm{pat}}$ (deg/day) &$r_{\rm{res}}$ (km)  \\
\midrule
$\ell-m=1$ \\
3 &2 &3175.4 - 90.1 &31704 + 612 \\
4 &3 &2661.6 - 72.2 &32959 + 608 \\
5 &4 &2329.7 - 60.7 &34494 + 611 \\
6 &5 &2094.4 - 52.7 &36031 + 615 \\
7 &6 &1918.8 - 46.6 &37482 + 619 \\
8 &7 &1783.2 - 42.0 &38818 + 621 \\
9 &8 &1675.8 - 38.4 &40033 + 623 \\
10 &9 &1589.0 - 35.5 &41133 + 624 \\
11 &10 &1517.7 - 33.1 &42127 + 624 \\
12 &11 &1458.4 - 31.2 &43021 + 623 \\
13 &12 &1408.6 - 29.5 &43823 + 622 \\
14 &13 &1366.5 - 28.1 &44541 + 621 \\
15 &14 &1330.5 - 26.9 &45184 + 619 \\
16 &15 &1299.3 - 25.9 &45767 + 617 \\
17 &16 &1271.9 - 24.9 &46301 + 615 \\
18 &17 &1247.4 - 24.1 &46796 + 612 \\
19 &18 &1225.3 - 23.4 &47259 + 610 \\
20 &19 &1205.2 - 22.7 &47695 + 607 \\
21 &20 &1186.7 - 22.0 &48107 + 605 \\
22 &21 &1169.7 - 21.5 &48499 + 602 \\
23 &22 &1153.9 - 20.9 &48872 + 599 \\
24 &23 &1139.3 - 20.4 &49228 + 597 \\
25 &24 &1125.6 - 20.0 &49569 + 594 \\
\midrule
$\ell-m=3$ \\
5 &2 &4231.9 - 121.5 &26202 + 511 \\
6 &3 &3193.5 - 87.8 &29202 + 545 \\
7 &4 &2649.1 - 70.0 &31670 + 568 \\
8 &5 &2309.8 - 58.9 &33760 + 584 \\
9 &6 &2076.8 - 51.2 &35559 + 596 \\
10 &7 &1906.7 - 45.7 &37125 + 604 \\
11 &8 &1777.1 - 41.4 &38500 + 609 \\
12 &9 &1675.3 - 38.1 &39711 + 613 \\
13 &10 &1593.5 - 35.4 &40782 + 615 \\
14 &11 &1526.6 - 33.2 &41731 + 616 \\
15 &12 &1471.0 - 31.4 &42576 + 616 \\
16 &13 &1424.0 - 29.9 &43335 + 616 \\
17 &14 &1383.5 - 28.5 &44025 + 615 \\
18 &15 &1348.1 - 27.3 &44658 + 613 \\
19 &16 &1316.8 - 26.3 &45245 + 611 \\
20 &17 &1288.7 - 25.3 &45792 + 610 \\
21 &18 &1263.4 - 24.5 &46305 + 608 \\
22 &19 &1240.5 - 23.7 &46787 + 606 \\
23 &20 &1219.5 - 23.0 &47243 + 603 \\
24 &21 &1200.2 - 22.4 &47675 + 601 \\
25 &22 &1182.4 - 21.8 &48085 + 599 \\
\midrule
$\ell-m=5$ \\
8 &3 &3538.6 - 98.1 &27279 + 514 \\
9 &4 &2879.0 - 76.8 &29967 + 543 \\
10 &5 &2478.6 - 63.9 &32214 + 564 \\
11 &6 &2209.5 - 55.2 &34125 + 579 \\
12 &7 &2016.1 - 49.0 &35773 + 590 \\
13 &8 &1870.8 - 44.3 &37205 + 598 \\
14 &9 &1757.9 - 40.6 &38459 + 603 \\
15 &10 &1667.7 - 37.7 &39564 + 607 \\
16 &11 &1594.0 - 35.3 &40549 + 609 \\
17 &12 &1532.3 - 33.3 &41435 + 610 \\
18 &13 &1479.7 - 31.5 &42241 + 610 \\
19 &14 &1434.3 - 30.0 &42980 + 610 \\
20 &15 &1394.5 - 28.7 &43663 + 609 \\
21 &16 &1359.3 - 27.6 &44297 + 608 \\
22 &17 &1327.9 - 26.5 &44888 + 607 \\
23 &18 &1299.6 - 25.6 &45443 + 606 \\
24 &19 &1274.0 - 24.7 &45965 + 604 \\
25 &20 &1250.6 - 24.0 &46457 + 602 \\
\bottomrule                          
\end{tabular}
\label{vert_thick_med}

\end{table}

\newpage
\begin{table}
\centering
\caption{Pattern frequencies and vertical resonance locations of different models
}
\renewcommand{\arraystretch}{0.5}
\begin{tabular}{@{}rrcc@{}}
\multicolumn{4}{c}{\textbf{Uranus Thin model}} \\
\toprule
$\ell$ &$m$ &$\Omega_{\rm{pat}}$ (deg/day) &$r_{\rm{res}}$ (km)  \\
\midrule
$\ell-m=1$ \\
3 &2 &3176.8 - 90.2 &31695 + 612 \\
4 &3 &2678.2 - 72.8 &32823 + 607 \\
5 &4 &2354.5 - 61.6 &34252 + 609 \\
6 &5 &2122.9 - 53.6 &35709 + 612 \\
7 &6 &1948.1 - 47.6 &37106 + 616 \\
8 &7 &1811.8 - 42.9 &38409 + 618 \\
9 &8 &1702.9 - 39.2 &39608 + 619 \\
10 &9 &1614.1 - 36.2 &40706 + 620 \\
11 &10 &1540.5 - 33.8 &41710 + 620 \\
12 &11 &1478.6 - 31.7 &42628 + 619 \\
13 &12 &1425.8 - 29.9 &43470 + 619 \\
14 &13 &1380.3 - 28.4 &44244 + 618 \\
15 &14 &1340.8 - 27.2 &44952 + 617 \\
16 &15 &1306.6 - 26.0 &45598 + 615 \\
17 &16 &1276.7 - 25.1 &46185 + 614 \\
18 &17 &1250.6 - 24.2 &46717 + 612 \\
19 &18 &1227.5 - 23.4 &47204 + 609 \\
20 &19 &1206.7 - 22.7 &47655 + 607 \\
21 &20 &1187.9 - 22.1 &48076 + 605 \\
22 &21 &1170.6 - 21.5 &48474 + 602 \\
23 &22 &1154.7 - 21.0 &48851 + 599 \\
24 &23 &1139.9 - 20.5 &49210 + 597 \\
25 &24 &1126.1 - 20.0 &49553 + 594 \\
\midrule
$\ell-m=3$ \\
5 &2 &4283.1 - 123.2 &25994 + 508 \\
6 &3 &3241.6 - 89.3 &28914 + 541 \\
7 &4 &2693.2 - 71.4 &31325 + 564 \\
8 &5 &2349.6 - 60.1 &33379 + 580 \\
9 &6 &2112.4 - 52.3 &35160 + 591 \\
10 &7 &1938.3 - 46.6 &36722 + 599 \\
11 &8 &1804.9 - 42.2 &38105 + 605 \\
12 &9 &1699.2 - 38.7 &39339 + 608 \\
13 &10 &1613.3 - 35.9 &40448 + 611 \\
14 &11 &1542.3 - 33.6 &41449 + 613 \\
15 &12 &1482.6 - 31.7 &42355 + 613 \\
16 &13 &1432.0 - 30.0 &43173 + 614 \\
17 &14 &1388.8 - 28.6 &43913 + 613 \\
18 &15 &1351.6 - 27.4 &44581 + 612 \\
19 &16 &1319.1 - 26.3 &45191 + 611 \\
20 &17 &1290.4 - 25.4 &45752 + 609 \\
21 &18 &1264.7 - 24.5 &46274 + 608 \\
22 &19 &1241.5 - 23.8 &46762 + 606 \\
23 &20 &1220.3 - 23.0 &47222 + 603 \\
24 &21 &1200.9 - 22.4 &47656 + 601 \\
25 &22 &1183.0 - 21.8 &48069 + 599 \\
\midrule
$\ell-m=5$ \\
8 &3 &3604.4 - 100.2 &26948 + 509 \\
9 &4 &2931.5 - 78.5 &29609 + 538 \\
10 &5 &2521.9 - 65.2 &31845 + 560 \\
11 &6 &2245.4 - 56.3 &33761 + 575 \\
12 &7 &2045.8 - 49.8 &35427 + 586 \\
13 &8 &1894.6 - 44.9 &36894 + 594 \\
14 &9 &1776.2 - 41.1 &38195 + 600 \\
15 &10 &1681.0 - 38.0 &39356 + 604 \\
16 &11 &1603.0 - 35.5 &40396 + 607 \\
17 &12 &1538.3 - 33.4 &41328 + 608 \\
18 &13 &1483.7 - 31.6 &42167 + 609 \\
19 &14 &1437.0 - 30.1 &42927 + 609 \\
20 &15 &1396.4 - 28.8 &43623 + 609 \\
21 &16 &1360.7 - 27.6 &44266 + 608 \\
22 &17 &1329.0 - 26.6 &44863 + 607 \\
23 &18 &1300.5 - 25.6 &45422 + 606 \\
24 &19 &1274.7 - 24.8 &45946 + 604 \\
25 &20 &1251.3 - 24.0 &46440 + 602 \\
\bottomrule                          
\end{tabular}
\begin{tabular}{@{}rrcc@{}}
\multicolumn{4}{c}{\textbf{Uranus Adiabatic model}} \\
\toprule
$\ell$ &$m$ &$\Omega_{\rm{pat}}$ (deg/day) &$r_{\rm{res}}$ (km)  \\
\midrule
$\ell-m=1$ \\
3 &2 &3051.9 - 86.3 &32550 + 626 \\
4 &3 &2487.2 - 66.9 &34478 + 631 \\
5 &4 &2170.8 - 56.0 &36153 + 634 \\
6 &5 &1965.1 - 48.9 &37591 + 635 \\
7 &6 &1818.9 - 43.8 &38839 + 635 \\
8 &7 &1708.7 - 40.0 &39935 + 634 \\
9 &8 &1622.1 - 37.0 &40911 + 633 \\
10 &9 &1551.7 - 34.6 &41789 + 631 \\
11 &10 &1493.1 - 32.5 &42587 + 629 \\
12 &11 &1443.4 - 30.8 &43318 + 627 \\
13 &12 &1400.5 - 29.4 &43992 + 625 \\
14 &13 &1363.0 - 28.1 &44616 + 622 \\
15 &14 &1329.9 - 26.9 &45199 + 620 \\
16 &15 &1300.4 - 25.9 &45743 + 617 \\
17 &16 &1273.8 - 25.0 &46255 + 615 \\
18 &17 &1249.8 - 24.2 &46737 + 612 \\
19 &18 &1227.9 - 23.4 &47193 + 610 \\
20 &19 &1207.8 - 22.8 &47625 + 607 \\
21 &20 &1189.4 - 22.1 &48035 + 605 \\
22 &21 &1172.3 - 21.6 &48426 + 602 \\
23 &22 &1156.5 - 21.0 &48799 + 600 \\
24 &23 &1141.8 - 20.5 &49155 + 597 \\
25 &24 &1128.0 - 20.0 &49496 + 595 \\
\midrule
$\ell-m=3$ \\
5 &2 &3924.0 - 112.0 &27549 + 534 \\
6 &3 &2985.3 - 81.4 &30539 + 566 \\
7 &4 &2504.9 - 65.7 &32871 + 586 \\
8 &5 &2209.7 - 56.0 &34769 + 599 \\
9 &6 &2008.3 - 49.3 &36362 + 607 \\
10 &7 &1861.0 - 44.4 &37730 + 612 \\
11 &8 &1747.9 - 40.7 &38926 + 615 \\
12 &9 &1657.9 - 37.7 &39987 + 617 \\
13 &10 &1584.4 - 35.2 &40938 + 617 \\
14 &11 &1522.8 - 33.2 &41800 + 617 \\
15 &12 &1470.5 - 31.4 &42587 + 617 \\
16 &13 &1425.2 - 29.9 &43310 + 616 \\
17 &14 &1385.7 - 28.6 &43978 + 615 \\
18 &15 &1350.8 - 27.4 &44599 + 613 \\
19 &16 &1319.7 - 26.4 &45179 + 611 \\
20 &17 &1291.7 - 25.4 &45722 + 610 \\
21 &18 &1266.4 - 24.6 &46232 + 608 \\
22 &19 &1243.4 - 23.8 &46714 + 606 \\
23 &20 &1222.3 - 23.1 &47170 + 604 \\
24 &21 &1203.0 - 22.5 &47601 + 601 \\
25 &22 &1185.1 - 21.9 &48012 + 599 \\
\midrule
$\ell-m=5$ \\
8 &3 &3378.7 - 93.3 &28129 + 528 \\
9 &4 &2780.7 - 74.0 &30667 + 555 \\
10 &5 &2417.7 - 62.2 &32751 + 573 \\
11 &6 &2172.6 - 54.2 &34509 + 585 \\
12 &7 &1995.1 - 48.5 &36023 + 594 \\
13 &8 &1860.2 - 44.0 &37347 + 600 \\
14 &9 &1753.7 - 40.5 &38521 + 604 \\
15 &10 &1667.3 - 37.7 &39572 + 607 \\
16 &11 &1595.5 - 35.3 &40522 + 609 \\
17 &12 &1534.9 - 33.4 &41388 + 610 \\
18 &13 &1482.9 - 31.6 &42182 + 610 \\
19 &14 &1437.6 - 30.1 &42914 + 610 \\
20 &15 &1397.9 - 28.8 &43592 + 609 \\
21 &16 &1362.7 - 27.7 &44224 + 608 \\
22 &17 &1331.1 - 26.6 &44815 + 607 \\
23 &18 &1302.8 - 25.7 &45369 + 606 \\
24 &19 &1277.0 - 24.8 &45891 + 604 \\
25 &20 &1253.6 - 24.1 &46383 + 602 \\
\bottomrule                          
\end{tabular}
\label{vert_thin_ad}

\end{table}

\newpage
\begin{table}
\centering
\caption{Pattern frequencies and vertical resonance locations of different models
}
\renewcommand{\arraystretch}{0.5}
\begin{tabular}{@{}rrcc@{}}
\multicolumn{4}{c}{\textbf{Uranus Shallow model}} \\
\toprule
$\ell$ &$m$ &$\Omega_{\rm{pat}}$ (deg/day) &$r_{\rm{res}}$ (km)  \\
\midrule
$\ell-m=1$ \\
3 &2 &3069.0 - 86.9 &32429 + 625 \\
4 &3 &2633.1 - 71.6 &33195 + 613 \\
5 &4 &2344.5 - 61.5 &34349 + 612 \\
6 &5 &2133.7 - 54.2 &35589 + 614 \\
7 &6 &1970.3 - 48.5 &36828 + 616 \\
8 &7 &1839.1 - 44.0 &38029 + 618 \\
9 &8 &1731.5 - 40.4 &39172 + 620 \\
10 &9 &1642.0 - 37.3 &40245 + 621 \\
11 &10 &1566.9 - 34.8 &41241 + 621 \\
12 &11 &1503.4 - 32.7 &42159 + 621 \\
13 &12 &1449.3 - 30.9 &43000 + 620 \\
14 &13 &1402.9 - 29.3 &43769 + 619 \\
15 &14 &1362.7 - 27.9 &44471 + 618 \\
16 &15 &1327.7 - 26.8 &45115 + 616 \\
17 &16 &1296.8 - 25.7 &45707 + 614 \\
18 &17 &1269.4 - 24.8 &46254 + 612 \\
19 &18 &1244.9 - 24.0 &46762 + 610 \\
20 &19 &1222.8 - 23.2 &47237 + 607 \\
21 &20 &1202.6 - 22.5 &47683 + 605 \\
22 &21 &1184.2 - 21.9 &48103 + 602 \\
23 &22 &1167.2 - 21.4 &48501 + 600 \\
24 &23 &1151.5 - 20.8 &48879 + 598 \\
25 &24 &1136.9 - 20.3 &49239 + 595 \\
\midrule
$\ell-m=3$ \\
5 &2 &4269.7 - 123.0 &26049 + 510 \\
6 &3 &3264.4 - 90.3 &28780 + 541 \\
7 &4 &2730.1 - 72.8 &31043 + 563 \\
8 &5 &2390.7 - 61.7 &32997 + 578 \\
9 &6 &2152.9 - 53.8 &34719 + 590 \\
10 &7 &1976.1 - 48.0 &36253 + 598 \\
11 &8 &1839.4 - 43.5 &37627 + 604 \\
12 &9 &1730.8 - 39.9 &38859 + 608 \\
13 &10 &1642.5 - 37.0 &39967 + 611 \\
14 &11 &1569.6 - 34.6 &40966 + 613 \\
15 &12 &1508.5 - 32.6 &41868 + 613 \\
16 &13 &1456.6 - 30.9 &42686 + 613 \\
17 &14 &1411.9 - 29.4 &43433 + 613 \\
18 &15 &1373.0 - 28.1 &44117 + 612 \\
19 &16 &1338.8 - 27.0 &44748 + 611 \\
20 &17 &1308.4 - 26.0 &45333 + 609 \\
21 &18 &1281.1 - 25.1 &45879 + 607 \\
22 &19 &1256.5 - 24.2 &46390 + 606 \\
23 &20 &1234.1 - 23.5 &46870 + 604 \\
24 &21 &1213.6 - 22.8 &47323 + 602 \\
25 &22 &1194.8 - 22.2 &47753 + 600 \\
\midrule
$\ell-m=5$ \\
8 &3 &3677.8 - 102.8 &26590 + 505 \\
9 &4 &2995.8 - 80.7 &29186 + 535 \\
10 &5 &2577.5 - 67.2 &31387 + 556 \\
11 &6 &2293.6 - 58.0 &33288 + 572 \\
12 &7 &2088.0 - 51.3 &34949 + 584 \\
13 &8 &1932.4 - 46.3 &36412 + 592 \\
14 &9 &1810.6 - 42.3 &37711 + 598 \\
15 &10 &1712.7 - 39.1 &38869 + 602 \\
16 &11 &1632.5 - 36.5 &39910 + 605 \\
17 &12 &1565.4 - 34.3 &40850 + 607 \\
18 &13 &1508.5 - 32.4 &41704 + 608 \\
19 &14 &1459.4 - 30.8 &42486 + 608 \\
20 &15 &1416.8 - 29.4 &43206 + 608 \\
21 &16 &1379.2 - 28.2 &43871 + 607 \\
22 &17 &1345.7 - 27.1 &44491 + 606 \\
23 &18 &1315.8 - 26.1 &45069 + 605 \\
24 &19 &1288.8 - 25.2 &45612 + 604 \\
25 &20 &1264.2 - 24.4 &46123 + 602 \\
\bottomrule                          
\end{tabular}
\begin{tabular}{@{}rrcc@{}}
\multicolumn{4}{c}{\textbf{Neptune model}} \\
\toprule
$\ell$ &$m$ &$\Omega_{\rm{pat}}$ (deg/day) &$r_{\rm{res}}$ (km)  \\
\midrule
$\ell-m=1$ \\
3 &2 &3211.4 - 63.9 &33238 + 447 \\
4 &3 &2659.9 - 50.8 &34830 + 449 \\
5 &4 &2326.8 - 42.8 &36470 + 454 \\
6 &5 &2103.6 - 37.5 &37953 + 457 \\
7 &6 &1943.3 - 33.7 &39264 + 460 \\
8 &7 &1822.2 - 30.8 &40423 + 462 \\
9 &8 &1726.8 - 28.6 &41458 + 463 \\
10 &9 &1649.4 - 26.7 &42391 + 464 \\
11 &10 &1584.9 - 25.2 &43241 + 464 \\
12 &11 &1530.1 - 23.9 &44021 + 464 \\
13 &12 &1482.9 - 22.8 &44741 + 464 \\
14 &13 &1441.6 - 21.8 &45411 + 463 \\
15 &14 &1405.1 - 20.9 &46036 + 463 \\
16 &15 &1372.5 - 20.2 &46623 + 462 \\
17 &16 &1343.1 - 19.5 &47175 + 461 \\
18 &17 &1316.6 - 18.9 &47697 + 460 \\
19 &18 &1292.3 - 18.3 &48191 + 459 \\
20 &19 &1270.1 - 17.8 &48660 + 459 \\
21 &20 &1249.7 - 17.3 &49107 + 458 \\
22 &21 &1230.8 - 16.8 &49534 + 457 \\
23 &22 &1213.2 - 16.4 &49941 + 455 \\
24 &23 &1196.8 - 16.0 &50332 + 454 \\
25 &24 &1181.5 - 15.7 &50706 + 453 \\
\midrule
$\ell-m=3$ \\
5 &2 &4256.1 - 85.7 &27568 + 374 \\
6 &3 &3230.4 - 62.5 &30610 + 400 \\
7 &4 &2702.6 - 50.6 &33013 + 417 \\
8 &5 &2377.7 - 43.2 &34983 + 429 \\
9 &6 &2155.7 - 38.1 &36645 + 437 \\
10 &7 &1993.3 - 34.4 &38079 + 443 \\
11 &8 &1868.6 - 31.5 &39336 + 448 \\
12 &9 &1769.4 - 29.2 &40455 + 451 \\
13 &10 &1688.1 - 27.3 &41462 + 453 \\
14 &11 &1620.2 - 25.8 &42377 + 455 \\
15 &12 &1562.3 - 24.4 &43215 + 456 \\
16 &13 &1512.2 - 23.3 &43987 + 457 \\
17 &14 &1468.5 - 22.3 &44703 + 457 \\
18 &15 &1429.8 - 21.4 &45370 + 457 \\
19 &16 &1395.3 - 20.6 &45994 + 457 \\
20 &17 &1364.2 - 19.8 &46580 + 457 \\
21 &18 &1336.2 - 19.2 &47132 + 457 \\
22 &19 &1310.6 - 18.6 &47655 + 456 \\
23 &20 &1287.2 - 18.1 &48150 + 455 \\
24 &21 &1265.6 - 17.6 &48620 + 455 \\
25 &22 &1245.7 - 17.1 &49068 + 454 \\
\midrule
$\ell-m=5$ \\
7 &2 &4980.5 - 101.1 &24839 + 340 \\
8 &3 &3673.9 - 71.9 &28103 + 371 \\
9 &4 &3013.5 - 57.1 &30708 + 393 \\
10 &5 &2612.4 - 48.1 &32860 + 409 \\
11 &6 &2341.5 - 42.0 &34684 + 420 \\
12 &7 &2145.3 - 37.6 &36262 + 429 \\
13 &8 &1995.9 - 34.2 &37648 + 435 \\
14 &9 &1878.1 - 31.5 &38881 + 440 \\
15 &10 &1782.3 - 29.3 &39990 + 444 \\
16 &11 &1702.8 - 27.5 &40996 + 447 \\
17 &12 &1635.6 - 26.0 &41915 + 449 \\
18 &13 &1577.8 - 24.7 &42760 + 451 \\
19 &14 &1527.6 - 23.5 &43543 + 452 \\
20 &15 &1483.5 - 22.5 &44270 + 453 \\
21 &16 &1444.3 - 21.6 &44949 + 453 \\
22 &17 &1409.2 - 20.8 &45585 + 454 \\
23 &18 &1377.6 - 20.1 &46183 + 454 \\
24 &19 &1348.9 - 19.4 &46748 + 454 \\
25 &20 &1322.8 - 18.8 &47282 + 453 \\
\bottomrule                          
\end{tabular}
\label{vert_shallow_Nep}

\end{table}

\clearpage

\section{Extended Tables of g-mode Frequencies and Resonance Locations}

In this appendix, we present all the calculated g-mode frequencies and resonance locations for all the models. 
Table \ref{U_g} displays the $\ell=m$, $n=1$ Lindblad resonance locations for Uranus.
Table \ref{N_g} shows the $\ell=m$, $n=1$ Lindblad resonance locations and the corotation resonance locations associated with them for Neptune.

\begin{table}[htp]
\addtolength{\tabcolsep}{-3pt}
\centering
\caption{$\ell=m$, $n=1$, g-mode pattern frequencies and Lindblad resonance locations of non-adiabatic Uranus models}
\renewcommand{\arraystretch}{0.5}
\begin{tabular}{@{}rcccccccc@{}}
 &\multicolumn{2}{c}{\textbf{Thick}} &\multicolumn{2}{c}{\textbf{Medium}} &\multicolumn{2}{c}{\textbf{Thin}} &\multicolumn{2}{c}{\textbf{Shallow}} \\
\toprule
$\ell$,$m$ &$\Omega_{\rm{pat}}$ (deg/day) &$r_{\rm{res}}$ (km) &$\Omega_{\rm{pat}}$ (deg/day) &$r_{\rm{res}}$ (km) &$\Omega_{\rm{pat}}$ (deg/day) &$r_{\rm{res}}$ (km) &$\Omega_{\rm{pat}}$ (deg/day) &$r_{\rm{res}}$ (km)  \\
\midrule
1  &1785.9 - 49.8 &56259 +1070 &1807.0 - 50.7 &55820 +1069 &1769.6 - 49.4 &56605 +1079 &1418.5 - 37.2 &65597 +1172 \\
2  &1723.2 - 42.2 &47569 + 793 &1745.5 - 43.2 &47163 + 794 &1717.1 - 42.2 &47682 + 798 &1426.3 - 31.9 &53959 + 819 \\
3  &1659.1 - 38.7 &45108 + 715 &1684.8 - 39.8 &44647 + 716 &1662.1 - 39.0 &45053 + 719 &1400.9 - 29.7 &50490 + 726 \\
4  &1602.1 - 36.2 &44230 + 679 &1632.0 - 37.5 &43688 + 681 &1614.4 - 36.9 &44005 + 683 &1375.4 - 28.3 &48962 + 684 \\
5  &1550.0 - 34.2 &44004 + 659 &1584.4 - 35.6 &43365 + 662 &1571.3 - 35.2 &43605 + 663 &1351.5 - 27.3 &48211 + 659 \\
6  &1501.1 - 32.4 &44118 + 646 &1540.3 - 34.0 &43367 + 649 &1531.2 - 33.7 &43539 + 651 &1328.6 - 26.4 &47857 + 644 \\
7  &1454.7 - 30.7 &44438 + 636 &1499.1 - 32.5 &43556 + 640 &1493.5 - 32.4 &43665 + 642 &1306.3 - 25.5 &47740 + 632 \\
8  &1410.1 - 29.1 &44897 + 628 &1460.5 - 31.1 &43859 + 634 &1457.8 - 31.1 &43912 + 635 &1284.3 - 24.8 &47780 + 624 \\
9  &1367.0 - 27.6 &45457 + 622 &1424.2 - 29.8 &44233 + 628 &1424.2 - 29.9 &44234 + 630 &1262.4 - 24.0 &47932 + 617 \\
10 &1325.5 - 26.2 &46091 + 616 &1390.1 - 28.6 &44653 + 623 &1392.4 - 28.8 &44604 + 626 &1240.6 - 23.3 &48168 + 611 \\
11 &1285.9 - 24.8 &46774 + 611 &1358.0 - 27.5 &45103 + 618 &1362.5 - 27.8 &45004 + 622 &1218.9 - 22.5 &48469 + 606 \\
12 &1248.5 - 23.6 &47482 + 606 &1327.7 - 26.4 &45576 + 614 &1334.4 - 26.8 &45422 + 618 &1197.3 - 21.8 &48823 + 601 \\
13 &1213.7 - 22.4 &48194 + 602 &1298.7 - 25.4 &46068 + 610 &1308.1 - 25.9 &45848 + 614 &1176.0 - 21.1 &49218 + 597 \\
14 &1181.5 - 21.4 &48898 + 598 &1271.0 - 24.4 &46577 + 606 &1283.3 - 25.0 &46278 + 610 &1154.9 - 20.4 &49644 + 592 \\
15 &1151.9 - 20.4 &49585 + 593 &1244.4 - 23.5 &47096 + 602 &1259.8 - 24.2 &46712 + 606 &1134.4 - 19.7 &50093 + 588 \\
16 &1124.6 - 19.5 &50250 + 589 &1219.2 - 22.7 &47619 + 599 &1237.4 - 23.4 &47152 + 603 &1114.4 - 19.1 &50557 + 584 \\
17 &1099.6 - 18.7 &50891 + 584 &1195.4 - 21.9 &48138 + 596 &1215.7 - 22.6 &47601 + 599 &1095.1 - 18.4 &51030 + 580 \\
18 &1076.6 - 17.9 &51508 + 579 &1173.0 - 21.1 &48646 + 593 &1194.7 - 21.9 &48057 + 596 &1076.6 - 17.8 &51506 + 576 \\
19 &1055.3 - 17.2 &52102 + 574 &1152.1 - 20.4 &49142 + 590 &1174.6 - 21.2 &48515 + 593 &1059.0 - 17.3 &51980 + 572 \\
20 &1035.6 - 16.6 &52672 + 570 &1132.6 - 19.8 &49622 + 587 &1155.4 - 20.6 &48968 + 590 &1042.2 - 16.7 &52449 + 568 \\
21 &1017.3 - 16.0 &53220 + 565 &1114.3 - 19.2 &50088 + 583 &1137.2 - 20.0 &49413 + 587 &1026.2 - 16.2 &52911 + 564 \\
22 &1000.3 - 15.4 &53748 + 560 &1097.2 - 18.6 &50538 + 580 &1120.0 - 19.4 &49848 + 584 &1011.1 - 15.7 &53363 + 560 \\
23 &984.4 - 14.9  &54255 + 555 &1081.1 - 18.1 &50973 + 577 &1103.9 - 18.9 &50270 + 581 &996.8 - 15.3  &53805 + 556 \\
24 &969.6 - 14.4  &54744 + 550 &1065.9 - 17.6 &51395 + 574 &1088.6 - 18.4 &50680 + 578 &983.2 - 14.8  &54236 + 552 \\
25 &955.7 - 14.0  &55214 + 545 &1051.7 - 17.2 &51803 + 571 &1074.2 - 17.9 &51078 + 575 &970.4 - 14.4  &54654 + 548 \\
\bottomrule   
\end{tabular}
\label{U_g}
\end{table}

\begin{table}[htp]
\centering
\caption{$\ell=m$, $n=1$, g-mode pattern frequencies and Lindblad resonance locations of our Neptune model}
\renewcommand{\arraystretch}{0.5}
\begin{tabular}{@{}rccc@{}}
\toprule
$\ell$,$m$ &$\Omega_{\rm{pat}}$ (deg/day) &$r_{\rm{res}}$ (km) &$r_{\rm{cor}}$ (km) \\
\midrule
1  &1808.1 - 34.7 &58959 + 766 &37170 + 481 \\
2  &1683.1 - 28.2 &51058 + 578 &38986 + 441 \\
3  &1566.6 - 24.6 &49518 + 526 &40893 + 433 \\
4  &1462.1 - 21.9 &49668 + 502 &42815 + 432 \\
5  &1369.0 - 19.7 &50502 + 489 &44732 + 432 \\
6  &1287.1 - 17.7 &51646 + 480 &46611 + 432 \\
7  &1215.6 - 16.1 &52917 + 472 &48417 + 431 \\
8  &1153.8 - 14.7 &54217 + 465 &50129 + 429 \\
9  &1100.5 - 13.5 &55495 + 458 &51736 + 426 \\
10 &1054.3 - 12.5 &56722 + 451 &53235 + 423 \\
11 &1014.1 - 11.6 &57888 + 444 &54630 + 418 \\
12 & 979.0 - 10.8 &58988 + 437 &55927 + 414 \\
13 & 948.2 - 10.1 &60022 + 429 &57132 + 408 \\
14 & 920.9 -  9.5 &60992 + 422 &58253 + 402 \\
15 & 896.7 -  8.9 &61902 + 415 &59299 + 396 \\
16 & 875.0 -  8.5 &62757 + 407 &60274 + 390 \\
17 & 855.5 -  8.0 &63560 + 400 &61186 + 385 \\
18 & 837.9 -  7.6 &64315 + 393 &62041 + 379 \\
19 & 821.9 -  7.3 &65027 + 386 &62843 + 373 \\
20 & 807.3 -  6.9 &65698 + 379 &63597 + 367 \\
21 & 793.9 -  6.6 &66331 + 373 &64308 + 361 \\
22 & 781.7 -  6.4 &66931 + 366 &64978 + 355 \\
23 & 770.4 -  6.1 &67498 + 360 &65612 + 349 \\
24 & 759.9 -  5.9 &68036 + 354 &66211 + 344 \\
25 & 750.2 -  5.7 &68547 + 348 &66779 + 339 \\
\bottomrule                          
\end{tabular}
\label{N_g}
\end{table}

\end{document}